\documentclass[aps,prl,twocolumn,showpacs,superscriptaddress,nofootinbib,groupedaddress]{revtex4-2}  
\usepackage{graphicx,color,braket}  
\usepackage{dcolumn}   
\usepackage{bm}        
\usepackage{amssymb}   
\usepackage{physics}
\usepackage{xparse}
\usepackage{amsmath}
\usepackage{hyperref}
\usepackage{cleveref}
\usepackage{latexsym, mathrsfs}
\usepackage[toc,page]{appendix}

\usepackage{subfigure}

\newcommand{\be}{\begin{equation}}
\newcommand{\ee}{\end{equation}}
\newcommand{\bc}{\begin{center}}
	\newcommand{\ec}{\end{center}}
\newcommand{\bea}{\begin{eqnarray}}
\newcommand{\eea}{\end{eqnarray}}
\newcommand{\bml}{\begin{subequations}}
	\newcommand{\eml}{\end{subequations}}
\newcommand{\bfig}{\begin{figure}}
	\newcommand{\efig}{\end{figure}}
\newcommand{\bmult}{\begin{multline}}
\newcommand{\emult}{\end{multline}}
\newcommand{\ag}{\alpha}
\newcommand{\bg}{\beta}

\newcommand{\bmat}{\begin{pmatrix}}
	\newcommand{\emat}{\end{pmatrix}}
\NewDocumentCommand{\tens}{t_}
 {%
  \IfBooleanTF{#1}
   {\tensop}
   {\otimes}%
 }
\NewDocumentCommand{\tensop}{m}
 {%
  \mathbin{\mathop{\otimes}\displaylimits_{#1}}%
}
\Crefname{appendix}{Appx.}{Appxs.}

\begin{document}



\title{\Large Indirect detection of Cosmological Constant from \\ 
	interacting open quantum system }
\author{Subhashish Banerjee${}^{1}$,~Sayantan Choudhury${}^{2,3,4*}$,\\ ~Satyaki Chowdhury${}^{3,4}$,~Rathindra Nath Das${}^{5}$,~Nitin Gupta${}^{6}$,\\
~Sudhakar Panda${}^{3,4}$,~Abinash Swain${}^{3,4}$} 
\thanks{{\it Corresponding author,}\\
			{{ E-mail:sayantan.choudhury@niser.ac.in}}}
	~~~~~~~~					
\affiliation{${}^{1}$Indian Institute of Technology Jodhpur, Jodhpur 342011, India.}
\affiliation{${}^{2}$Quantum Gravity and Unified Theory and Theoretical Cosmology Group, Max Planck Institute
for Gravitational Physics (Albert Einstein Institute), Am M$\ddot{u}$hlenberg 1, 14476 Potsdam-Golm,
Germany.}
\affiliation{${}^{3}$National Institute of Science Education and Research, Jatni, Bhubaneswar, Odisha - 752050, India.}
\affiliation{${}^{4}$Homi Bhabha National Institute, Training School Complex, Anushakti Nagar, Mumbai - 400085,
India.}
\affiliation{${}^{5}$Department of Physics, Indian Institute of Technology Bombay, Powai, Mumbai - 400076, India.}
\affiliation{${}^{6}$Department of Physical Sciences,
Indian Institute of Science Education \& Research Mohali,
Manauli PO 140306 Punjab India.}
\begin{abstract}
We study the indirect detection of {\it Cosmological Constant} from an {\it open quantum system} of 
interacting spins, weakly interacting with a thermal bath, a massless scalar field minimally coupled with the static de Sitter background, by computing the spectroscopic shifts. By assuming pairwise 
interaction between spins, we construct  
states using a generalisation of the superposition principle.  
The corresponding spectroscopic shifts, caused by the effective Hamiltonian of the system due to Casimir Polder interaction, 
are seen to play a crucial role in predicting a very tiny value of the Cosmological Constant, 
in the static patch of de Sitter space,  which is consistent with the observed value from the Planck measurements of the cosmic microwave background (CMB) anisotropies.
\end{abstract}

\maketitle 


 In recent times the study of the quantum systems that are interacting with their surroundings has acquired a lot of attention in different fields ranging from condensed matter \cite{weiss,Hu,Rammer,Kamenev}, quantum information \cite{Banerjee:2019b}, subatomic physics \cite{Banerjee:2014vga,Dixit:2019lsg,Naikoo:2019eec,Naikoo:2018amb,Naikoo:2018vug,Alok:2015iua}, quantum dissipative systems \cite{Chakrabarty:2018dov}, holography \cite{Chakrabarty:2019aeu, Jana:2020vyx} to cosmology \cite{Akhtar:2019qdn, Maldacena:2012xp, Kanno:2014lma, Kanno:2014ifa, Kanno:2014bma, Kanno:2015lja, Kanno:2016gas, Kanno:2016qcc, Kanno:2017dci, Albrecht:2018prr, Kanno:2018cuk, Kanno:2019gqw, Kanno:2020lpi, Choudhury:2018fpj, Choudhury:2017qyl, Choudhury:2017bou, Martin:2012pea, Martin:2015qta, Martin:2016tbd, Martin:2016nrr, Martin:2017zxs, Martin:2018zbe, Green:2020whw, Yu:2011eq, Benatti:2004ee, Tian:2014jha, Huang:2017yjt, Choudhury:2018rjl, Kiefer:2008ku, Choudhury:2018bcf, Breuer:2003Ox, Banerjee:2010An, Banerjee:2011QI, Banerjee:2019b}. Here our interest is the study of the curvature of the static patch of de Sitter space as well as the Cosmological Constant from
the spectroscopic Lamb shift \cite{Zhou:2010nb, Tian:2016uwp, Bhattacherjee:2019eml}. Wave equation and Hawking radiation in de Sitter space-time have been studied in \cite{Polarski, Polarski:1989iu}. The system under consideration is an open quantum system of $N$ 
interacting spins which are weakly coupled to their environment, modelled by a massless scalar field minimally coupled to static patch of de Sitter space-time. We are interested to see the effect of the curvature of the static patch of de Sitter space-time as well as the Cosmological Constant on the 
states of the system and the Lamb shift when the number of spins become very large in the thermodynamic limit. 
One can design such a thought experimental condensed matter analogue gravity \cite{Barcelo:2005fc, Visser:2001fe} set up of measuring spectroscopic shift in an open quantum system in a quantum laboratory to get a proper estimation of the curvature of the static patch of de Sitter space as well as the Cosmological Constant without recourse to any cosmological observation. This is the main highlight of this work, where our claim is that, without doing any cosmological observation one can measure the value of the Cosmological Constant from quantum spectroscopy of open systems. We show from our analysis that the obtained value of the Cosmological Constant is perfectly consistent with the present day observed central value of the Cosmological Constant, $\Lambda_{\rm observed}\sim 2.89\times 10^{-122}$ in the Planckian unit \cite{Aghanim:2018eyx} and is completely independent of the number of entangled spins. 
Computational details, and some relevant material, are expounded in a number of Appendices. A detailed calculation of the $N$-point Wightman function is given in Appendix A and its Hilbert transformation in Appendix B .  We have also added Appendix C  and Appendix D, which shows the detailed construction of quantum mechanical states by providing explicit examples of $2$ and $3$ spin systems.  Next, in Appendix E we have presented the generalised version of the previously discussed Appendix C and D with an arbitrary $N$ number of spins.  We also discuss the thermodynamic large $N$ limiting situation and the flat space limit of the spectroscopic shifts in the next Appendices F and G.  Finally,  in Appendix H, we provide a detailed derivation of the bath scalar field Hamiltonian in the static patch of de Sitter space.

The open quantum set up can be described by the following Hamiltonian:
\begin{equation}
H_{\text{T}} = H_{\text{S}}\otimes {I}_{\text{2,B}} + {I}_{\text{2,S}}\otimes H_{\text{B}} + H_{\text{I}},
\end{equation}
where $H_{\text{S}}$,  $H_{\text{B}}$ and $H_{\text{I}}$ respectively describes the Hamiltonian of the spin system, bath and the interaction between them.  The strength of the interaction between the system and the thermal bath is usually characterized by a coupling parameter,  the value of which considered to be large for an exactly solvable non-perturbative quantum systems (strongly coupled theory) or taken to be very small for the systems which can be solved only perturbatively (weakly coupled theory).  Here it is important to note that,  the magnitude of this coupling parameter of interaction is actually compared to some fundamental physical parameter which is appearing in the system Hamiltonian.  If the strength of the interaction parameter is small compared to the characteristic parameter appearing in the system Hamiltonian,  then the system is solvable in the perturbative regime of the theory.  In the contrary case,  if the model is exactly solvable to all orders of perturbation theory then only one can take the strong coupling  approximation.  Hence the applicability of weak coupling or strong coupling limits vary for the specific choices of the system and the interaction Hamiltonian and the analysis is completely model dependent in the context of open quantum systems.  Also ${I}_{\text{2,S}}$ and ${I}_{\text{2,B}}$ are the identity operators for the system and bath, respectively. We choose our spin Hamiltonian in such a way that the individual Pauli matrices are oriented arbitrarily in space. In the present context, the $N$ spin system Hamiltonian is described by:
\begin{equation} \label{system}
	H_{S} = \frac{\omega}{2} \sum_{\delta=1}^{N}\sum^{3}_{i=1} n^\delta_i.\sigma^{\delta}_{i},
	\end{equation}
where $n^\delta_i$ represent the unit vectors along any arbitrary $i(=1,2,3)$-th direction for $\delta=1,\cdots,N$. Also, $\sigma^{\delta}_i$, $(i=1,2,3)$, are the three usual Pauli matrices for each particle characterized by the particle number index $\delta$.  It is important to note that,  the quantity $\omega$ represents the renormalized energy level for $N$ identical spins ~\footnote{Here $\omega$ appearing in the system Hamiltonian is not the same as the natural frequency of the $N$ identical spins,  which is actually represented by the symbol $\omega_0$ in the present context of discussion.  It can be understood the significance of the background space-time in this context for which instead of $\omega_0$ the factor $\omega$ is appearing along with the renormalized correction factors.  In the flat space limit, the $\omega$ appearing in the system Hamiltonian will be just replaced by $\omega_0$,  whereas for in any curved space-time (in our case for the static patch of de Sitter space-time),  the $\omega$ has two major contributions, one coming from the fundamental frequency $\omega_{0}$, which is actually the flat space limiting result and the other significant contribution coming from the Hilbert transformation of the Fourier transformed Wightman functions which we have defined in the appendix of the paper.  This additional part actually captures the information regarding the background curved space-time in this discussion.}, given by:
\be
	\displaystyle \omega=\omega_0+i\times [\mathcal{K}^{(\delta\delta)}(-\omega_{0})-\mathcal{K}^{(\delta\delta)}(\omega_{0})]~~~~~~\delta=1,\cdots,N \ee	
Here $\mathcal{K}^{\delta \delta}(\pm \omega_{0})$ for $\delta \in \{1,2,\cdots ,N\}$ are Hilbert transformations of Wightman function computed from the probe massless scalar field, which we have defined explicitly in appendix section of this paper. 

The free rescaled scalar field, minimally coupled with the static de Sitter background is considered as the bath, and is described by the following Hamiltonian: 
	\begin{widetext}\begin{eqnarray}\label{wqw}
	&&H_{ B}=\int^{\infty}_{0} dr~\int ^{\pi}_{0}d\theta~\int^{2\pi}_{0}d\phi~\left[\frac{\Pi^2_{\Phi}}{2}+\frac{r^2\sin^2\theta}{2}\left\{r^2~(\partial_{r}\Phi)^2+ \displaystyle \frac{1}{\displaystyle\left(1-\frac{r^2}{\alpha^2}\right)}\left((\partial_{\theta}\Phi)^2+\frac{1}{\sin^2\theta}(\partial_{\phi}\Phi)^2\right)\right\}\right].~~~~~~~
	\end{eqnarray}
	\end{widetext}
The details of the Hamiltonian has been provided in Appendix H.
Here, $\Pi_{\Phi}$ represents the  momentum canonically conjugate to the scalar field $\Phi(x)$ in the static de Sitter patch. As a choice of background classical geometry, here we have considered the static de Sitter patch, as our prime objective is to implement the present methodology to the real world cosmological observation. The static de Sitter metric (which we will define later) contains the Cosmological Constant term explicitly which is one of the prime measurable quantities at late time scale (mostly at the present day) in Cosmology. Using this analogue gravity thought experiment performed with $N$ spins our objective is to measure the value of Cosmological Constant at present day from the spectroscopic shift formula indirectly.  The choice of De-Sitter space as the background geometry comes from the assumption of identifying our universe with an exponentially flat expanding universe. The proof concerning the validity of the approximation is beyond the scope of this work. For this purpose we have only taken the observed value of Cosmological Constant to check the consistency of our finding from this methodology. Not only the numerical value of the Cosmological Constant, but also the curvature of static patch of de Sitter space can be further constrained using the present methodology.  The interaction between the $N$ identical spin system and the thermal bath plays a crucial role in the dynamics of open quantum system.  For the model being considered, the interaction between the system of $N$ entangled spins and the bath is given by: 
\begin{eqnarray}
H_{I}= \mu \sum_{\delta=1}^{N}\sum^{3}_{i=1} (n^\delta_i.\sigma^{\delta}_{i})\Phi(x^{\delta}),~~~~
\end{eqnarray}
where the parameter $\mu$ represents the coupling strength of the interaction between the $N$ identical spin system with the (environment) thermal bath,  where it is modelled by a massless scalar field.  Since in the presence of the above mentioned type of interaction the problem is not exactly solvable for this reason we need treat such interaction perturbatively and this can only be implemented if we consider the coupling strength is sufficiently small enough compared to the renormalized energy scale $\omega$ i.e.  $\mu\ll \omega$.  Also, it is important to note that in the interaction Hamiltonian we have restricted upto quadratic contribution. Any higher order non-linear interactions are avoided for the sake of simplicity, but for a generalised case one can include such contributions in the present analysis. 

Motivated by the work of Fleck and Tanas \cite{Ficek},  where the space of a two-atom system, in the presence of damping due to dipole-dipole interaction was constructed, we construct below the space spanned by normalized $N$ spin states appropriate to the Hamiltonian under consideration. They are four states of interest, namely,  the ground state $|G\rangle$,  the excited state $|E\rangle$,  the symmetric as well as the anti-symmetric entangled states, $|S\rangle$ and $|A\rangle$ respectively and are given by:
\begin{eqnarray}\label{states}
&&|{G}\rangle \propto  \sum_{\delta,\eta=1,\delta<\eta}^{N} |{g_\delta}\rangle \otimes |{g_\eta}\rangle ,~~~\\
&&|{E}\rangle \propto  \sum_{\delta,\eta=1,\delta<\eta}^{N}|{e_\delta}\rangle \otimes |{e_\eta}\rangle ,~~\nonumber\\
&&|{S}\rangle, |{A}\rangle \propto  \sum_{\delta,\eta=1,\delta<\eta}^{N}\frac{1}{\sqrt{2}}\left( |{e_\delta}\rangle \otimes |{g_\eta}\rangle \pm  |{g_\delta}\rangle \otimes |{e_\eta}\rangle\right)~,~~~~~\nonumber
\end{eqnarray}
where $|{g_\delta}\rangle,|{e_\eta}\rangle \forall \delta,\eta=1,\cdots,N$ are the eigen vectors for individual spin corresponding to ground (lower energy) state and excited (higher energy) state.  The structure of the states reveals that the concept of quantum entanglement between the interacting spins can be realized from the symmetric and the antisymmetric states.  Hence these are physically relevant.  We would like to clarify that, in this work we have actually used $N$ interacting spin system.  However the interaction is pairwise.  Thus the total Hilbert space to construct the wave function has been taken into account and the total Hilbert space has ${}^NC_2$ number of interactions.  For this reason,  the combinatorial factor ${}^NC_2$ appears in the normalization factor of the wave function.  In an $N$ pairwise interacting spin system, there exists ${}^N C_2$ such cases of interacting pairs.  Without using the assumption of pairwise interaction, it is really difficult to generalize the symmetric and antisymmetric states for higher spin systems where interaction exists among all the spins simultaneously.  For this reason we define the proportionality constant of the normalization factor as:
\bea {\cal N}_{\rm norm}=\frac{1}{\sqrt{{}^N C_2}}=\sqrt{\frac{2(N-2)!}{N!}}.\eea
The normalization constant has been fixed by taking the inner products between elements of the direct product space with the restriction that the inner product only acts between elements belonging to the same Hilbert space of the open quantum system under consideration.  Some examples of the construction of states for $2$ and $3$ spin case  are provided in Appendix C.
 
 At the starting point we assume separable initial conditions, i.e., the total density matrix $\rho_{T}$  at the initial time scale $\tau=\tau_0$ factorizes 
as, $\rho_{T}(\tau_0)=\rho_{S}(\tau_0) \otimes \rho_{B}(\tau_0),$ where $\rho_{S}(\tau_0) $ and $\rho_{B}(\tau_0) $ constitute the system and bath  density matrices at initial time $\tau=\tau_0$, respectively.
As the system evolves with time, it starts interacting with its surrounding which we have treated as a thermal bath modelled by massless scalar field placed in the static de Sitter background. Since we are interested in the dynamics of our system of interest (sub system), made by the $N$ spins, we consider its reduced density matrix  by taking partial trace over the thermal bath, i.e.,
$\rho_{ S}(\tau) = {\rm Tr}_{B} [\rho_{T}(\tau)]$.
Though the total system plus bath joint evolution is unitary, the reduced dynamics of the system of interest is not. The non-unitary dissipative time evolution of the reduced density matrix of the sub system in the weak coupling limit can be described by the GKSL (Gorini Kossakowski Sudarshan Lindblad) master equation \cite{Akhtar:2019qdn}, 
$\partial_{\tau} \rho_{S}(\tau)= -i[H_{ \rm eff},\rho_{ S}(\tau)] + \mathcal{L}[\rho_{S}(\tau)]$,
where $\mathcal{L}[\rho_{S}(\tau)]$ is the Lindbladian operator which captures the effects of quantum dissipation and non-unitarity. The effective Hamiltonian, for the present model, is 
$H_{\text{eff}}=H_{\text{S}}+H_{\text{LS}}$,
where $H_{\text{LS}}(\tau)$ is the Lamb shift Hamiltonian given by:
\begin{equation}
H_{\text{LS}} = - \frac{i}{2}\sum_{\delta,\eta=1}^{N}\sum_{i,j=1}^{3}H^{(\delta \eta)}_{ij}(n^\delta_i.\sigma^{\delta}_{i})(n^\eta_j.\sigma^{\eta}_{j}). 
\end{equation}
The assumption of pairwise interaction between the spins can be implemented in terms of the Pauli operators as, $\sigma^{\delta}_{i}=\sigma_i \otimes I_{2} $ (for first spin of the interacting pair), $\sigma^{\delta}_{i}=I_{2} \otimes \sigma_i$ (for second spin of the interacting pair) and $\sigma^{\delta}_{i}=I_{2} \otimes I_{2}$ (for all other non-interacting spins). To bring out the clarity of notation, we illustrate using a 3 spin interacting system here.\footnote{For a simple three spin system, considering pairwise interaction we have three possibilities; spin 1 and spin 2 interacting (spin 3 is non-interacting), spin 2 and spin 3 interacting (spin 1 non-interacting), spin 1 and spin 3 interacting (spin 2 non-interacting). Consider the case when spin 1 and spin 3 interacts and spin 2 doesn't participate in the interaction. In this case, spin 1 is represented by $\sigma^{1 }_{i}=\sigma_i \otimes I_{2} $ (upper index denotes the spin number and $i$ (=1,2,3) denotes the direction cosines and $\sigma_{i}$ are usual Pauli matrices), spin 2 is represented by $\sigma^{2}_{i}=I_{2} \otimes I_{2}$ (for spin 2) and spin 3 is represented by $\sigma^{3}_{i}=I_{2} \otimes \sigma_i$ (for spin 3).}. This way of representing the spins agrees with the way the states have been constructed and ensures that operations like taking expectation value of the Lamb shift Hamiltonian with the constructed states is well defined.  This part necessarily captures the interaction between the $N$ identical spins as it has contribution from all possibilities together.  For condensed matter system ($e.g.$ in Ising model),  one usually considers nearest neighbour interactions to study the quantum correlations and the phenomena of phase transition. The above mentioned Hamiltonian mimics the role of the well known Ising model Hamiltonian and physically represents a  spin-spin interaction Hamiltonian in Heisenberg spin chain model,  which arises from the interaction among the $N$ identical spins and the external massless scalar field describing the thermal bath.  The specific terminology {\it Lamb shift} was first used in hydrogen-like atomic system where in the lowest order approximation in the fine structure constant energy level shift was determined from an effective Hamiltonian.  Following the same,  in consistence with refs.\cite{Tian:2016uwp,Tian:2014jha},  we have used the terminology {\it Lamb shift} to describe the spectroscopic shift of the effective Hamiltonian describing spin-spin self interaction.

In the Lamb shift the time dependent coefficient matrix $H^{(\delta \eta)}_{ij}(\tau)$ can be obtained from the Hilbert transform of the 
Wightman function, (refer to Appendix A and B for the details of the computation of the Wightman function) which is computed in the static de Sitter patch, described by the following 4D infinitesimal line element \cite{Spradlin:2001pw}:
\begin{widetext}
 \begin{eqnarray}
 \label{metric}
 ds^{2}=\left(1-\frac{r^{2}}{\alpha^{2}}\right)dt^{2}-\frac{1}{\displaystyle \left(1-\frac{r^{2}}{\alpha^{2}}\right)}dr^{2}-r^{2}d\Omega_2~~~~~{\rm where}~~~~~~d\Omega_2=\left(d\theta^2+r^2\sin^2\theta~ d\phi^2\right)~~{\rm with}~~\alpha=\sqrt{\frac{3}{\Lambda}}.
 \end{eqnarray}
 \end{widetext}
 where $\Lambda>0$ is the 4D Cosmological Constant in Static de Sitter patch. We use the Schwinger Keldysh technique to determine the entries of each 
 Wightman functions \footnote{The effect of dS spacetime enters through 
  the Wightman functions},  which are basically two point functions in quantum field theory at finite temperature. Consequently, the diagonal entries (auto-correlations) of the 
Wightman function are calculated as \cite{Tian:2016uwp}:
\begin{eqnarray}
\label{auto}
G^{\alpha \alpha}(x,x') &=& G^{\beta \beta}(x,x')=- \frac{1}{16\pi^{2}k^{2}\sinh^2f(\Delta\tau,k)},~~~~~~   
\end{eqnarray}
where we define, $f(\Delta\tau,k)=\left(\Delta \tau/2k-i\epsilon\right)$ and $\epsilon$ is an infinitesimal contour deformation parameter.  Also the off-diagonal (cross-correlation) components of the 
Wightman function can be computed as\cite{Tian:2016uwp}:
\begin{eqnarray}
\label{cross}
G^{\alpha\beta}(x,x')  &=& G^{\beta \alpha}(x,x')\nonumber\\
&=&\frac{-(16\pi^{2}k^{2})^{-1}}{\left\{\sinh^2f(\Delta\tau,k)-\frac{r^{2}}{k^{2}}\sin[2](\frac{\Delta \theta}{2})\right\}}.~~~~~~
\end{eqnarray}
 Here the parameter $k$ can be expressed as,
\bea k=\sqrt{g_{00}}\alpha=\sqrt{\alpha^2-r^2}=\sqrt{3/\Lambda-r^2}>0 \eea
Further, the curvature of the static de Sitter patch can be expressed in terms of the Ricci scalar term, given by,
$R=12/\alpha^2.$ This directly implies that one can probe the Cosmological Constant from the static de Sitter patch using the spectroscopic shift. The shifts for identical $N$ entangled spins can be physically interpreted as the energy shift obtained for each individual spin immersed in a thermal bath, described by the temperature, 
\bea T=\frac{1}{\beta}=\frac{1}{2\pi k}=\sqrt{T^2_{\rm GH}+T^2_{\rm Unruh}},\eea (with Planck's constant $\hbar=1$ and Boltzmann constant $k_{B}=1$) where the {\it Gibbons-Hawking} and {\it Unruh} temperature are defined as \cite{Tian:2016uwp,Bhattacherjee:2019eml}.,  
	\bea T_{\rm GH}=\frac{1}{2\pi \alpha},~T_{\rm Unruh}=\frac{a}{2\pi},~~{\rm with}~a=\frac{r}{\alpha^2}\left(1-\frac{r^2}{\alpha^2}\right)^{-1/2}.
	\eea 
Here the temperature of the bath $T$ can also be interpreted as the equilibrium temperature which can be obtained by solving the GKSL master equation for the thermal density matrix in the large time limit. Initially when the non-unitary system evolves with time it goes out-of-equilibrium and if we wait for long enough time, it is expected that the system will reach thermal equilibrium. The $N$ dependency comes in the states, in the matrix $H^{\delta\eta}_{ij}$ and the direction cosines of the alignment of each spin. The generic Lamb shifts are given by, 
$\delta E_{\Psi} = \langle \Psi| H_{LS} | \Psi \rangle,$
where $|\Psi\rangle$ is any possible entangled state. Here the spectral shifts for the $N$ spins are derived as:
\begin{eqnarray}
\frac{\delta E_{Y}^{N}}{2\Gamma_{1;{\cal DC}}^N}=\frac{\delta E_{S}^{N}}{\Gamma_{2;{\cal DC}}^N}=-\frac{\delta E_{A}^{N} }{\Gamma_{3;{\cal DC}}^N}=-{\cal F}(L,k,\omega_0) /{\cal N}^2_{\rm norm}, ~~~~
	\end{eqnarray} 
where $Y$ represents the ground and the excited states and $S$ and $A$ 
symmetric and antisymmetric states, respectively. Here, $\Gamma^N_{i;{\cal DC}}~\forall~i=1,2,3$ represent the direction cosine dependent angular factor which appears due to the fact that we have considered any arbitrary orientation of $N$ number of identical spins. These angular factors become extremely complicated to write for any arbitrary number of $N$ spins. Explicit expressions of the angular factors for 2 and 3 spin cases are provided in Appendix D. Because of this fact it is also expected that as we approach the large $N$ limit we get extremely complicated expressions. For all the spectral shifts we get an overall common factor of ${\cal N}^{-2}_{\rm norm}={}^NC_2=N!/2(N-2)!$ which is originating from the expectation value of the Lamb Shift Hamiltonian. Here we introduce a spectral function ${\cal F}(L,k,\omega_0)$, given by,
\begin{eqnarray}\label{sdsd}
 {\cal F}(L,k,\omega_0)&=&{\cal E}(L,k)\cos\left(2\omega_0 k\sinh^{-1}\left(L/2k\right)\right),~~~
 \end{eqnarray}
where,  we define:
\bea {\cal E}(L,k)=\mu^2/(8\pi L\sqrt{1+(L/2k)^2}). \eea 
In this context, $L$ represents the euclidean distance between any interacting pair of spins, and is
$L=2r\sin(\Delta \theta/2)$, where $\Delta \theta $ represents the angular separation, which we have assumed to be the same for all the interacting pairs of spins.  With respect to the length scales $L$ and $k$, we have two asymptotic solutions $L\gg k$ and $L\ll k$. In $L\gg k$ limit, the effect of the curvature of the static patch of de Sitter space is dominant and from the previously mentioned metric as stated in eq.~(\ref{metric}) at the horizon $r=\alpha$ we have ~$k=0$.  As a result, at horizon the limit $L\gg k$ corresponds to $L\gg 0$,  which implies the effect of the curvature of the static patch of de Sitter space can be probed exactly at the horizon of the metric stated in eq.~(\ref{metric}). This computation can be similarly performed for a near horizon region where one can take $r=\alpha-\Delta$.  Therefore, for a near horizon region one can write,  $k=\sqrt{\alpha^2-(\alpha-\Delta)^2}=\sqrt{(2\alpha-\Delta)\Delta}$.  In the near horizon case,  we can write $L\gg k=\sqrt{(2\alpha-\Delta)\Delta}$ and this again implies the fact that the effect of the curvature of the static patch of de Sitter space can be probed at the near horizon region as well. In the other limit $L\ll k$,  the curvature of the static patch of de Sitter space is not distinguishable and one can treat the space-time as a flat one which is described by the following metric:
\begin{equation}
\label{flatmetric}
ds^2=dt^2-(dr^2+r^2 d\Omega^2).
\end{equation} 
Since the horizon $r=\alpha$ corresponds to $k=0$, $L \ll k$ translates to $L \ll 0,$ $i.e.,$ it requires Euclidean distance to be negative which is impossible. This means, in this region, where the spacetime geometry is described by a flat metric, the notion of horizon does not exist.
The behaviour of the spectral function in these asymptotic limits can be seen to be:
\begin{widetext}
\begingroup
\large
\begin{eqnarray}\label{efg}
{\cal F}(L,k,\omega_0)=
 \left\{
     \begin{array}{lr}
  \displaystyle \frac{\mu^2 k}{4\pi L^2}\cos\left(2\omega_0 k\ln\left(L/2k\right)\right), ~~~~~~~~~~~~~~~~~~~~~~& \text{$L>>k$}\\ \\
\displaystyle  \frac{\mu^2}{8\pi L}\cos\left(\omega_0 L\right). ~~~~~~~~~~~~~~~~~~~~~~~~~& \text{$L<< k$}
    \end{array}
   \right.~~~~~
\end{eqnarray}
\endgroup
\end{widetext}
For a realistic situation we take the large $N$ limit, using the {\it Stirling-Gosper} approximation \cite{Gosper}, as a result of which the normalization factor can be written as:
\begin{widetext}
\begin{eqnarray}
&&{\cal N}_{\rm norm}~ \underrightarrow{\rm Large~N}~\widehat{{\cal N}_{\rm norm}} \approx\sqrt{2}\left(1-\frac{2}{\left(N+\frac{1}{6}\right)}\right)^{1/4}\left(\frac{N}{e}\right)^{-\displaystyle\frac{N}{2}}\left(\frac{N-2}{e}\right)^{N/2-1}\sqrt{\frac{\displaystyle1-\frac{2}{\left(N+\frac{1}{12}\right)}}{\displaystyle\left(1-\frac{2}{N}\right)}}.~~~~ \end{eqnarray}
\end{widetext}
Here we explicitly use the following result for factorial at large $N$ limit:
\bea N!\sim \sqrt{\left(2N+\frac{1}{3}\right)\pi }\left(\frac{N}{e}\right)^{N}\left(1+\frac{1}{12N}\right).\eea 
In general when we are talking about large number of degrees of freedom, instead of taking direct $N\rightarrow \infty$ limit in the combinatorial formula appearing in $N_{norm}$,  Stirling's approximation is very useful to correctly and almost accurately estimate the factorials.  This approximation allows us to take $N!$ in the large $N$ limit more accurately. 
\footnote{
Note:\quad The similar type of large $N$ approach is frequently used in the context of QFT to study the behaviour of the theory as a $O(1/N)$-th order perturbation theory,  which helps to understand the behaviour of the theory not only at $N\rightarrow \infty$,  but also in the intermediate finite value of $N$ where weak coupling behaviour holds good.  Actually,  within the framework of QFT strong coupling behaviour is very difficult to study,  hence an usual approach consists of translating the original theory in the weak coupling regime and solving by taking into account $O(1/N)$-th order perturbation theory.  Here the factor $1/N$ is treated as the perturbation parameter.  The good part of this approximation technique is that it helps to understand the intermediate weak coupling behaviour in terms of Feynman amplitudes and in the perturbative level those diagrams are computable and solvable.  To demonstrate the power of such techniques we want to cite an example,  a Chern-Simons Matter Theory where the general approach is to solve the theory in $O(1/N)$-th order perturbation theory to see the behaviour of the theory in weak coupling regime \cite{Choudhury:2018iwf,Choudhury:2017tax,Jain:2013gza,Giombi:2011kc}. } More discussions on this large $N$  approximation are made in Appendix F. 

Considering this fact carefully shifts can be approximately derived for large $N$ limiting case as :
\begin{eqnarray}
\frac{\widehat{\delta E_{Y}^{N}}}{2\Gamma_{1;{\cal DC}}^N}=\frac{\widehat{\delta E_{S}^{N}}}{\Gamma_{2;{\cal DC}}^N}=-\frac{\widehat{\delta E_{A}^{N}} }{\Gamma_{3;{\cal DC}}^N}=-{\cal F}(L,k,\omega_0) /\widehat{{\cal N}_{\rm norm}}^2. ~~~~~~~~
	\end{eqnarray} 

\begin{figure*}[htb]
  \includegraphics[width=17cm,height=8cm]{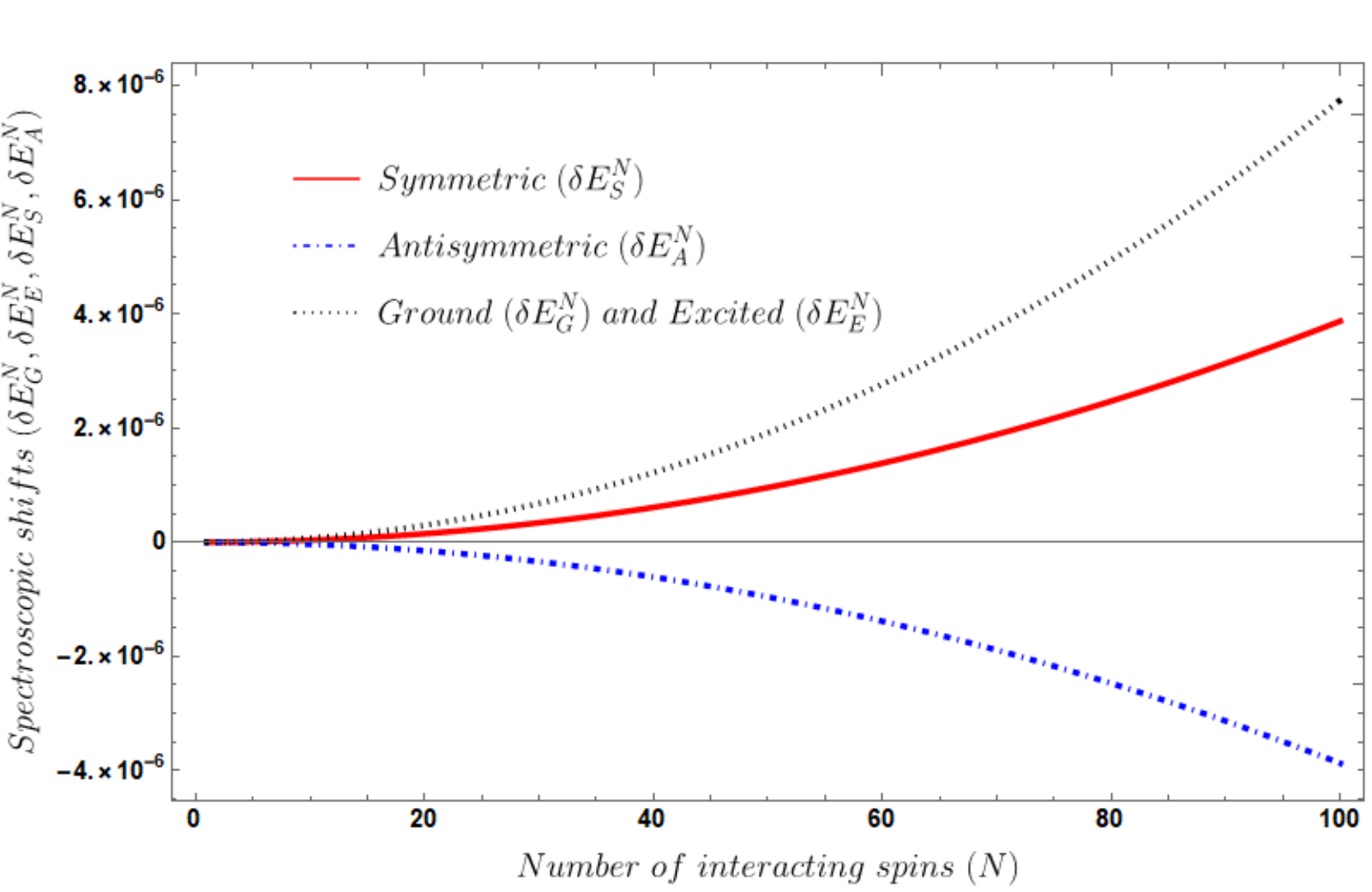}
  \caption{Behaviour of the spectroscopic shifts with the number of spins when the number of spins are small.  Here we fix the coupling parameter $\mu=10^{31}$,  Euclidean distance $L=10^{63}$,  Cosmological Constant $\Lambda=2\times 10^{-122}$ and natural frequency of the identical spins $\omega_0=10^{32}$.   Here the plot is made in such a way that $L\gg k$ and $\mu/\omega_0<1$ both are explicitly satisfied.  Here $k=\sqrt{3/\Lambda}$.  For all the parameters we have used here natural and Planckian unit system,  which implies $\hbar=c=1$ and $M_{\rm pl}=1$.}
  \label{fig:1}
\end{figure*}

\begin{figure*}[htb]
	\includegraphics[width=17cm,height=8cm]{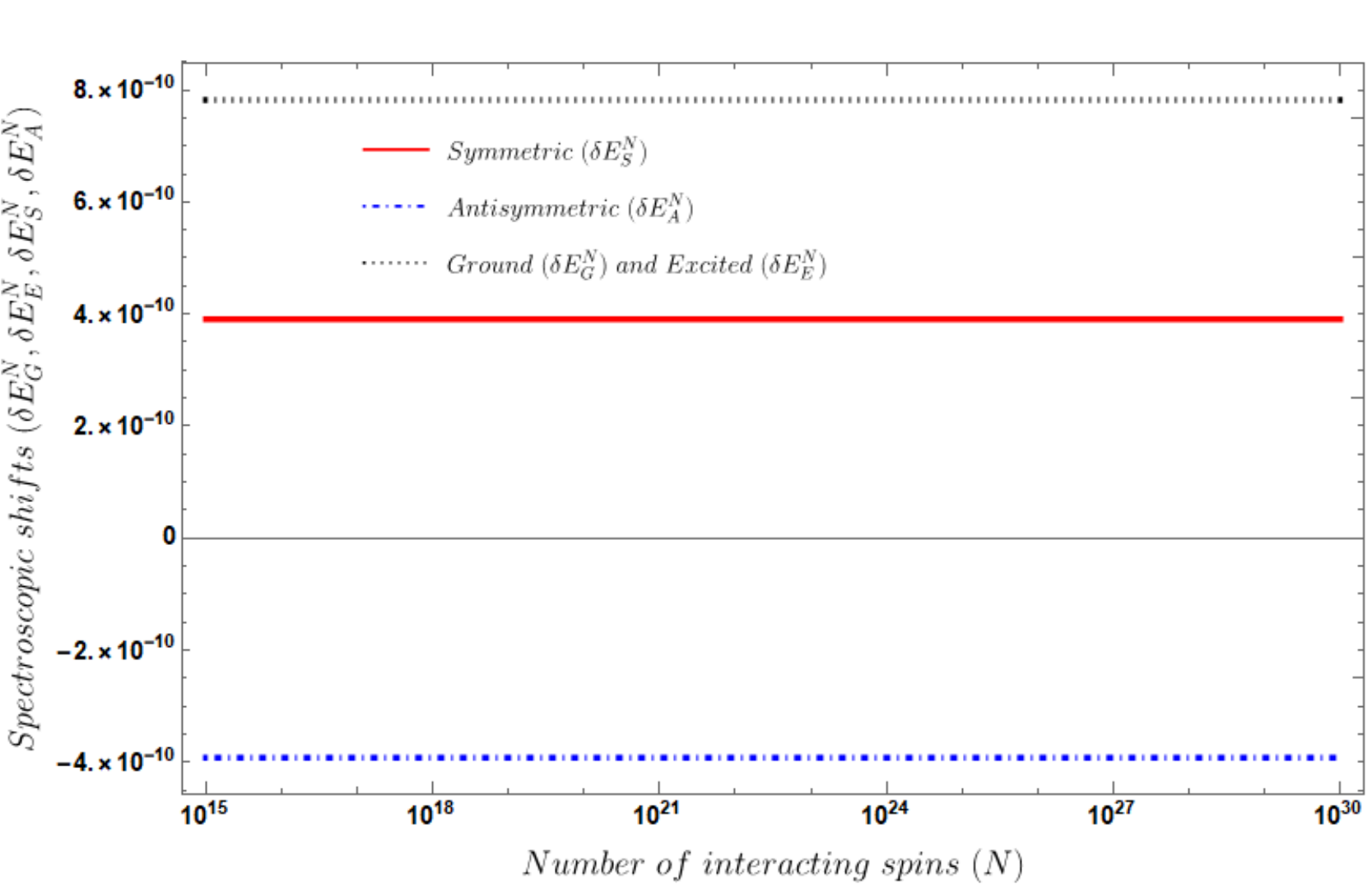}
	\caption{Behaviour of the spectroscopic shifts with the number of spins are very large.  Here we fix the coupling parameter $\mu=10^{31}$,  Euclidean distance $L=10^{63}$,  Cosmological Constant $\Lambda=2\times 10^{-122}$ and natural frequency of the identical spins $\omega_0=10^{32}$.   Here the plot is made in such a way that $L\gg k$ and $\mu/\omega_0<1$ both are explicitly satisfied.  Here $k=\sqrt{3/\Lambda}$.  For all the parameters we have used here natural and Planckian unit system,  which implies $\hbar=c=1$ and $M_{\rm pl}=1$.}
	\label{fig:1new}
\end{figure*}

\begin{figure*}[htb]
		\includegraphics[width=17cm,height=7.8cm]{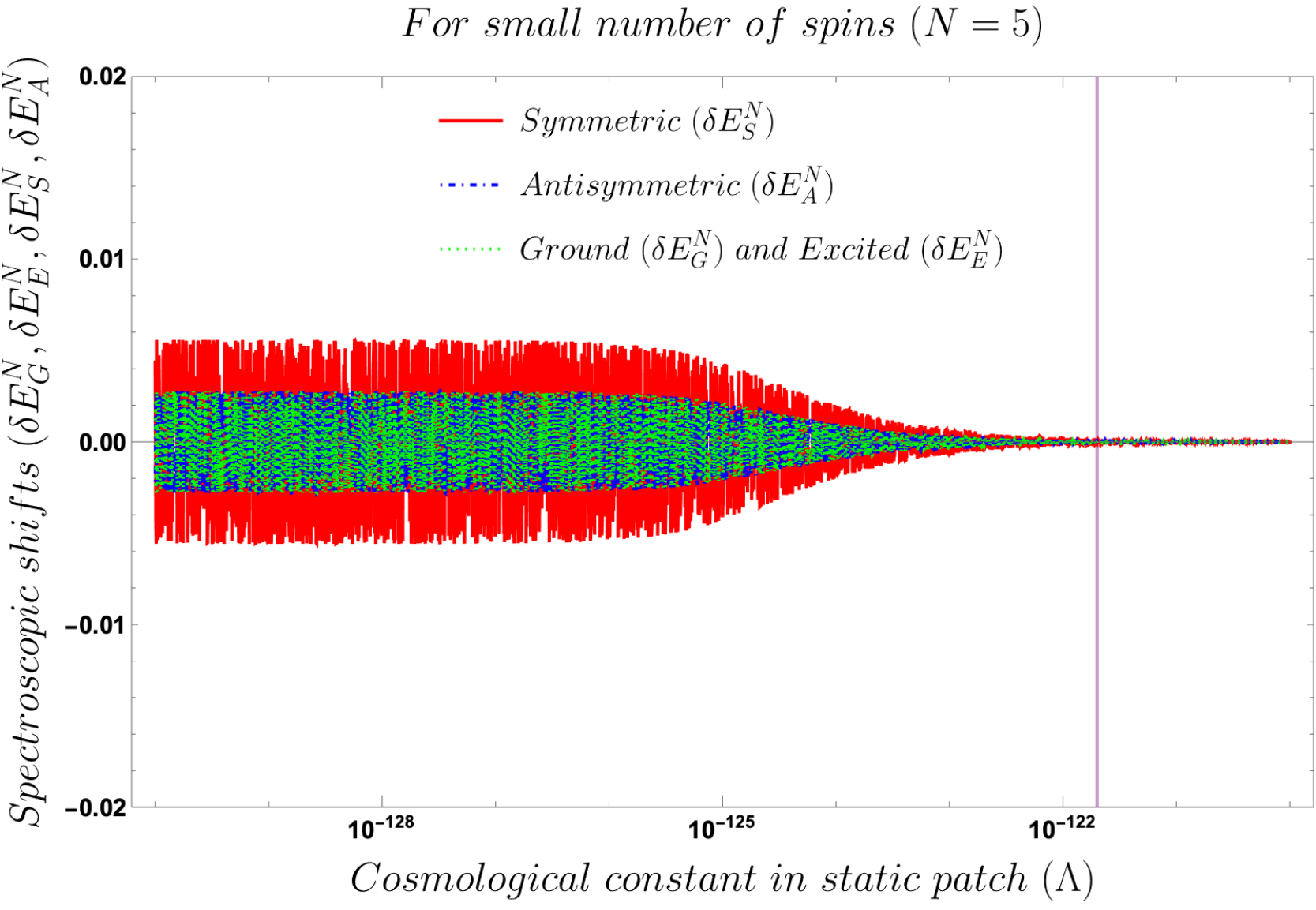}
		\caption{Behaviour of the spectroscopic shifts with the Cosmological Constant for small number of spins ($N=5$). Here we fix the coupling parameter $\mu=10^{31}$,  Euclidean distance $L=10^{63}$ and natural frequency of the identical spins $\omega_0=10^{32}$.  Here the plot is made in such a way that both $L\gg k$ and $\mu/\omega_0<1$  are explicitly satisfied.  For $L\ll k$ no pattern is observed in this context.  Here $k=\sqrt{3/\Lambda}$.  For all the parameters we have used here natural and Planckian unit system,  which implies $\hbar=c=1$ and $M_{\rm pl}=1$. }
		\label{fig:2}
	\end{figure*}
\begin{figure*}[htb]
 \includegraphics[width=17cm, height=7.8cm]{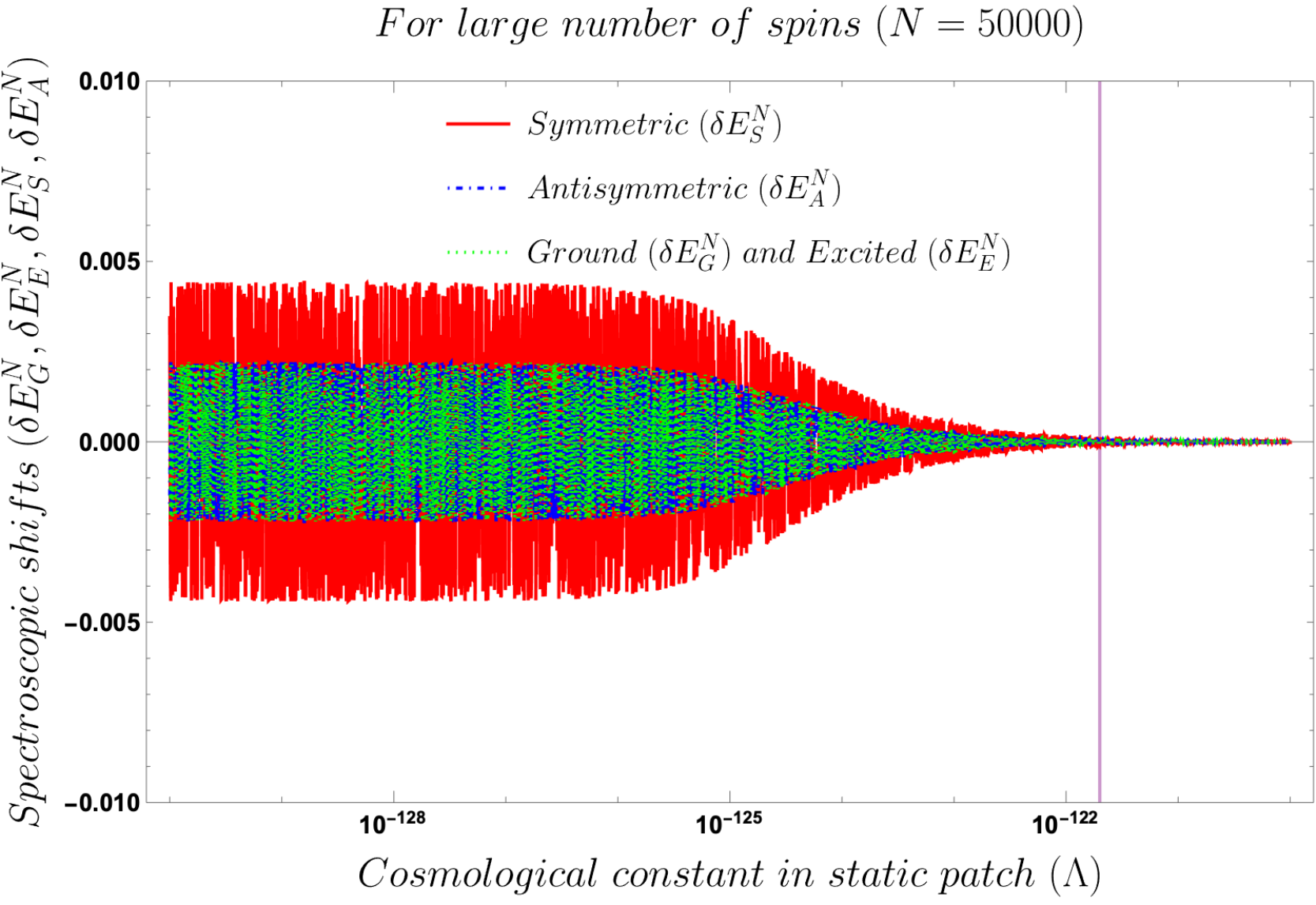}
 \caption{Behaviour of the spectroscopic shifts with the Cosmological Constant for large number of spins ($N=50000$). Here we fix the coupling parameter $\mu=10^{31}$,  Euclidean distance $L=10^{63}$ and natural frequency of the identical spins $\omega_0=10^{32}$.  Here the plot is made in such a way that both $L\gg k$ and $\mu/\omega_0<1$ are explicitly satisfied.  For $L\ll k$ no pattern is observed in this context.  Here $k=\sqrt{3/\Lambda}$.  For all the parameters we have used here natural and Planckian unit system,  which implies $\hbar=c=1$ and $M_{\rm pl}=1$.}
 \label{fig:3}
\end{figure*}
\begin{figure*}[htb]
		\includegraphics[width=17cm,height=7.8cm]{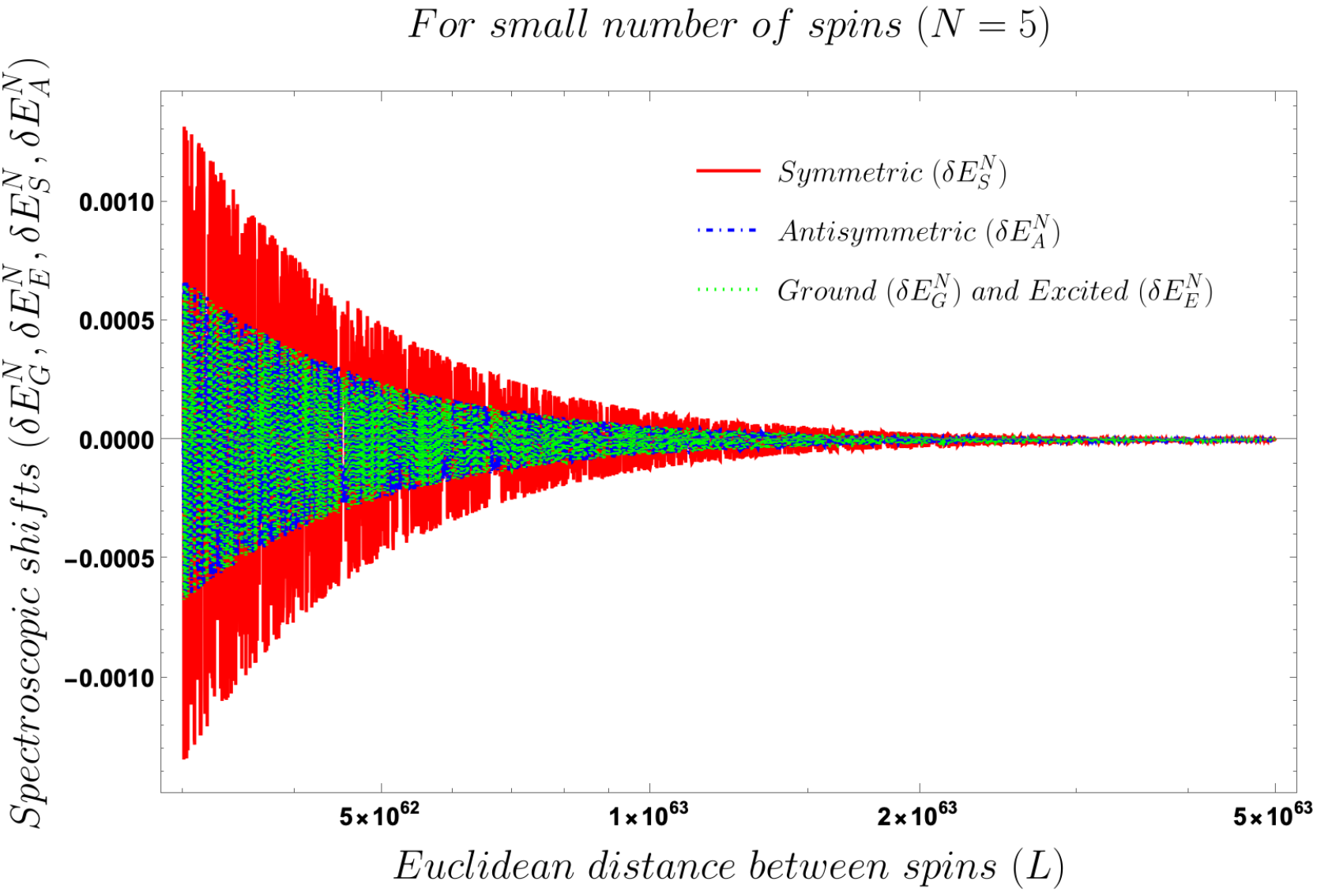}
		\caption{Behaviour of the spectroscopic shifts with the Euclidean distance for small number of spins ($N=5$). Here we fix the coupling parameter $\mu=10^{31}$,  Cosmological Constant $\Lambda=2\times 10^{-122}$ and natural frequency of the identical spins $\omega_0=10^{32}$ for which  $\mu/\omega_0<1$ approximation works perfectly well in this context.  Here the plot is made in such a way that $L\gg k$ is explicitly satisfied.  For $L\ll k$ no pattern is observed in this context.  Here $k=\sqrt{3/\Lambda}$.  For all the parameters we have used here natural and Planckian unit system,  which implies $\hbar=c=1$ and $M_{\rm pl}=1$.}  
		\label{fig:4}
	\end{figure*}
\begin{figure*}[htb]
		\includegraphics[width=17cm, height=7.8cm]{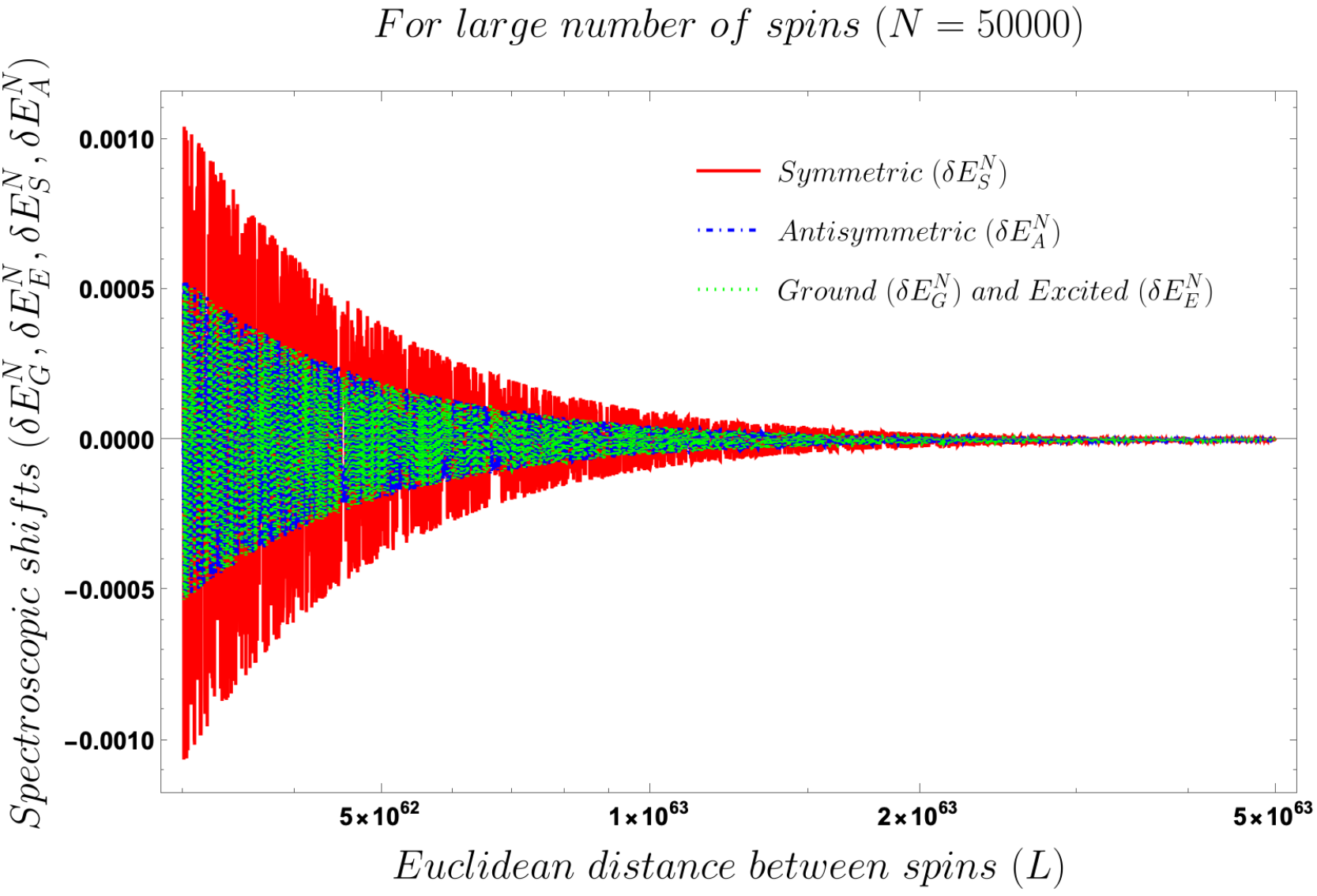}
		\caption{Behaviour of the spectroscopic shifts with the Euclidean distance for large number of spins ($N=50000$). Here we fix the coupling parameter $\mu=10^{31}$,  Cosmological Constant $\Lambda=2\times 10^{-122}$ and natural frequency of the identical spins $\omega_0=10^{32}$ for which  $\mu/\omega_0<1$ approximation works perfectly well in this context. Here the plot is made in such a way that $L\gg k$ is explicitly satisfied.  For $L\ll k$ no pattern is observed in this context.  Here $k=\sqrt{3/\Lambda}$.  For all the parameters we have used here natural and Planckian unit system,  which implies $\hbar=c=1$ and $M_{\rm pl}=1$.}  
		\label{fig:5}
	\end{figure*}
In the large $N$ limit, behaviour of ${\cal F}(L,k,\omega_0)$ remains unchanged, as the euclidean distance $L$, inverse of the curvature parameter $k$ and the frequency $\omega_0$ of the $N$ number of identical spins are not controlled by $N$. Also, for large $N$ the normalization factor asymptotically saturates to $\sqrt{2}\left(1+1/2N\right)$. 


In fig.~(\ref{fig:1}) and fig.~(\ref{fig:1new}), the behaviour of the spectroscopic shifts with the number of spins are depicted for two different scenarios, i.e., when the number of spins are small and large,  respectively.  From the first plot it can be seen that the magnitude of the spectroscopic shifts increases monotonically with the number of spins. This is justified because small number of spins is not a realistic situation and the shifts are sensitive to the number of spins present in the system. However, when the number of spins becomes extremely large,  close to Avagadro's number, the shifts becomes independent of $N$ and a constant value in the spectroscopic shifts is observed. This independence of the spectroscopic shifts of the number of spins agrees well with the physical intuition of any measurement being insensitive to fluctuations in the number of spins in the thermodynamic limit; otherwise it won't be a good measurement. Now  such variations in the value of $N$ will not effect the prediction in the value of the Cosmological Constant from the spectroscopic studies.  This crucial issue is explicitly discussed below.  Here it is important to note that, the scaling in these plots is different because of the presence of ${\cal F}(L,k,\omega_0)$ which we have fixed by fixing the coupling parameter $\mu=10^{31}$,  Euclidean distance $L=10^{63}$, Cosmological Constant  $\Lambda=2\times 10^{-122}$ and natural frequency of the identical spins $\omega_0=10^{32}$.  For all the relevant parameters (coupling parameter, Euclidean distance, cosmological constant, natural frequency of the spins and spectroscopic shifts) we have used here natural and Planckian unit system,  which implies $\hbar=c=1$ and $M_{\rm pl}=1$ and this is very useful for the estimation purpose.  Here the plots are made in such a way that $L\gg k$ and $\mu/\omega_0<1$ both are explicitly satisfied.  

In fig.~(\ref{fig:2}) and fig.~(\ref{fig:3}),  the behaviour of the spectroscopic shifts with respect to the Cosmological Constant are depicted, for a given small $N$ ($N=5$) and large $N$ ($N=50000$),  respectively.  For both of these plots we have fixed the coupling parameter $\mu=10^{31}$,  Euclidean distance $L=10^{63}$ and natural frequency of the identical spins $\omega_0=10^{32}$.  Here the plots are made in such a way that $L\gg k$ ($L\sqrt{\Lambda}\gg \sqrt{3}$) and $\mu/\omega_0<1$ both are explicitly satisfied.  In this context it is important to note that for both the plots in the region $L\ll k$ ($L\sqrt{\Lambda}\ll \sqrt{3}$) no pattern is observed which is consistent with the flat space time limiting behaviour.  More precisely,  in the $L\ll k$ region no effects of the curvature of the static patch of de Sitter space time and Cosmological Constant appear.  For this reason if we really want to capture effects and outcome of the curvature of the static patch of de Sitter space time and measure Cosmological Constant out of this analysis it is physically appropriate to look into the region which satisfy $L\gg k$.  From the behaviour of both the plots,  it is quite clear that the nature of spectroscopic shifts when studied with respect to variation of the Cosmological constant is almost independent of the number of interacting pair of spins present in the system.  In these plots, the vertical line corresponds to the value of the Cosmological Constant $\Lambda\sim 2.89\times 10^{-122}$ in the Planckian unit,  which is probed by the recent Planck data \cite{Aghanim:2018eyx} in cosmological observation~\footnote{Important note:~First of all,  we need to mention here that Cosmological Constant $\Lambda$ is not the direct observable in cosmological observation.   The Cosmological Constant $\Lambda$ can be expressed in terms of the product of two quantities i.e.  $\Lambda=3H^{2}_0\Omega_{\Lambda}$,  where $H_0$ rand $\Omega_{\Lambda}$ represent the present day value of the Hubble parameter and dimensionless density parameter due to sole contribution of Cosmological Constant.  Here,  $H_0$ rand $\Omega_{\Lambda}$ are probed by cosmological observations,  not directly the value of $\Lambda$.    If we know the values of both of them then by multiplying them one can give an estimate for the value of Cosmological Constant.  Using the recent Planck data \cite{Aghanim:2018eyx} for cosmological observation and parameter estimation it was found that $H_0=68.19\pm 0.67$ km~${\rm sec}^{-2}$~${\rm Mpc}^{-1}$ and $\Omega_{\Lambda}=0.6889\pm 0.0056$.  Now using these values in the $\Lambda=3H^{2}_0\Omega_{\Lambda}$ relation one can estimate that the consistent value of the Cosmological Constant in the Planckian unit scale turns out to be of the order of $\Lambda=2.89\times 10^{-122}$. }.  From both of these plots it is clearly observe that if we are able to measure the spectroscopic shifts from our model within the detector sensitivity range $-0.005\leq \delta E^{N}_{S},\delta E^{N}_{A},\delta E^{N}_{G},\delta E^{N}_{E}\leq 0.005$ \footnote{Note: The specific values used for the various parameters of the system are physically consistent to match the Cosmological Observation but might not be fully realistic for a typical experimental setup.  Also it is important to note that the values of the spectroscopic shifts obtained and represented in the plots are the outcome of the thought experiment we have designed. However, we believe that the prediction of the value of the Cosmological Constant from the setup will be unchanged if an actual experiment can be performed with fully realistic values of the model parameters. However,  we believe that the prediction of the value of the Cosmological Constant from the setup will be unchanged if an actual experiment can be performed with fully realistic values of the model parameters.} then it is possible to achieve the value of the Cosmological Constant $O(10^{-122})$ when fluctuations \footnote{Note: Here it is important to note that,  the rapid fluctuation in the spectroscopic shifts for the region $L\gg k$ is appearing from the cosine factor $\cos\left(2\omega_0 k\ln\left(L/2k\right)\right)$ and controlled by the factors,  $k=\sqrt{3/\Lambda}$,  $\omega_0$ and the ratio $L/k$.  See equation \ref{efg} for more details.  This fluctuation is basically the signature of the underlying quantum features of the open system considered in the static patch of the de Sitter background geometry. } are smaller.  The other values of the Cosmological Constant within the range $O(10^{-130}-10^{-123})$ might not be completely redundant,  but due to having large fluctuations in the spectroscopic shift pattern the estimation might not be strongly reliable for Cosmological Constant from this type of computation.  So for this reason when the fluctuations become small and the corresponding amplitude of the shifts are smaller (though lying within above mentioned sensitivity window of the detector which detects the shifts) one can reliably do the search for the observationally consistent value of the Cosmological Constant within the range $O(10^{-123}-10^{-121})$.  This range of the Cosmological Constant is very crucial for the parameter space that we have chosen to achieve this specific pattern of the shifts for the coupling parameter $\mu$,  Euclidean distance $L$ and the natural frequency of the all identical spins $\omega_0$.  We have chosen $L={\cal O}(10^{63})$,  $\mu={\cal O}(10^{31})$ and $\omega_0={\cal O}(10^{32})$,  which helps us to perfectly maintain the $L\gg k$ and $\mu/\omega_0<1$.   For other parameter space,  where $L\ll k$ and $\mu/\omega_0<1$ if we plot the shifts then we will get parallel lines and no effects of the curvature of the space time is reflected on that because this is actually representing the flat space result,  in which we are not at all interested in.   Our main motivation is extract the information of the curvature of the space time and the Cosmological Constant from the spectroscopic shift which is only possible to achieve strictly in the limit $L\gg k$.  One can further say instead of predicting the observationally consistent value of Cosmological Constant,  our analysis is able to predict a tiny window of Cosmological Constant $O(10^{-123}-10^{-121})$ within which one can realise the effect of curvature of the static de Sitter space and also fortunately the observationally consistent estimated value of the Cosmological Constant will lie in that window.

From fig.~(\ref{fig:4}) and fig.~(\ref{fig:5}) we again observe that the behaviour of the spectroscopic shifts is independent of the number of interacting pairs of spins $N$ (small or large),  for the Cosmological Constant fixed at the cosmologically estimated observed value $O(10^{-122})$.  In both of these plots are made to show the behaviour of all the spectroscopic shifts with respect to the Euclidean distance among all the identical spins for the specific model we are considering.  In both of the plots we have varied the Euclidean distance among the identical spins within the small window $O(10^{62}-10^{64})$ within which the limit $L\gg k$ is satisfied.  In principle this plot can be done for $L\ll k$ region also,  but that will going to give us only huge fluctuations having extremely large amplitude from which one cannot predict anything at all at the end of this analysis.  This oscillations having high frequency is appearing due to having the cosine term which dominates in this limit in the expressions for the spectroscopic shifts. We have already shown in fig.~(\ref{fig:2}) and fig.~(\ref{fig:3}) that the observationally consistent value of the Cosmological Constant can be achieved by having $L=O(10^{63})$ where the amplitude of the fluctuations in the shifts are small.  Here also in fig.~(\ref{fig:4}) and fig.~(\ref{fig:5}) we have observed the same feature.  Fluctuations in the shifts are small for $L=O(10^{63}-10^{64})$and this region can be reliably used to maintain $L\gg k$ and achieve the desired value of Cosmological Constant.   One might wonder the utility of doing the analysis with small number of interacting spins,  when generally considerations of cosmological studies involve large number of degrees of freedom. Though we are mainly interested in working with the thermodynamic limit which can be achieved through large number of degrees of freedom, the analysis with small number of degrees of freedom brings out the independence of the obtained result on the number of interacting spins. 
 
There emerge two natural length scales in the problem: one from the system, i.e., $L$ which is the Euclidean distance between two consecutive neighbouring spins and another from the bath $k$, which is related to the curvature and the cosmological constant.  An interplay between these two scales leads to rich dynamical consequences.  For $L \ll k$, one can find an inertial frame where the laws of Minkowski space-time are valid and the present shifts reduce to the flat space limit result. A more detailed discussion on this issue is given in Appendix G.  For $L \gg k$, the curvature of the static patch of de Sitter space-time dominates and plays a non-trivial role in spectral shifts. Here, the spectral shifts vary as $L^{-2}$ and depend explicitly on $k$. These are related to the Cosmological Constant $\Lambda$ and can be further linked to the equilibrium temperature of the bath.  For this reason we
will focus on the distances $L\gg k$ to have a non-trivial
effect.  For $L \ll k$, the spectral shifts vary as $L^{-1}$ and are independent of $k$ or $\Lambda$ for which the shifts should be essentially the same, as obtained in Minkowski case.  Presence of $k$ in the shifts for $L\gg k$ confirms the presence of  $\Lambda$ in the de Sitter static patch, which is of course, not present in the other limit, i.e., $L\ll k$. We have found, $\Lambda \sim {\cal O}(10^{-122})$ in the Planckian unit; this corresponds to almost constant shifts, which is consistent with the observed value, $\Lambda_{\rm observed}\sim 2.89\times 10^{-122}$ in Planckian unit \cite{Aghanim:2018eyx}.

Finally,  our theoretical analysis predicts a range of the value of cosmological constant which is not dependent on the number of interacting spins and also consistent with the observed bound on the Cosmological Constant obtained from other  observational probes \cite{Garnavich:1997nb, Schmidt:1998ys, Riess:1998cb, Garnavich:1998th, Perlmutter:1996ds, Perlmutter:1997zf, Perlmutter:1998np, Padmanabhan:2002ji, Perlmutter:2003kf}. From this analysis one can comment on a range in which the value of the observable can lie,  an issue of obviously interest.  Here,  we observe that the Cosmological Constant predicted from the Lamb shift spectroscopy can have a value between $O(10^{-123}-10^{-121})$,  which is consistent with the observed central value,    
${\cal O}(10^{-122})$.  Hence,  we can say that our theoretical analysis predicts the Cosmological Constant within a certain window.  
Generally,  if an analysis using CMB data is carried out,  predicting a particular value of the Cosmological Constant is possible along-with having cosmic variance from CMB,  but it is difficult to achieve from a theoretical calculation.  However, Fisher information techniques could  be useful in this regard \cite{Choudhury:2020dpf}. 
The analogue gravity thought process set-up discussed here helps us to probe such an important tiny fine tuned number from a theoretical perspective. 

In conclusion, we have studied indirect detection mechanism of observationally relevant Cosmological Constant from the shifts obtained from a realistic model of open system consisting of interacting $N$ spins. For this purpose, we have utilized the superposition principle along with equal Euclidean distance between all the spins.  In this work we found that-\begin{itemize}

\item  The shifts are not sensitive to the number of spins $N$,  

\item a correct prediction of a range of the observationally consistent Cosmological Constant \cite{Aghanim:2018eyx} can be made in the region where the Euclidean distance between any value of the spins is large compared to the length scale $k$ (i.e., $L\gg k$), irrespective of the number of the interacting spins.  In this region one can explicitly realize the effect of the curvature of the static patch of the de Sitter space which we have taken as our background metric for this computation and

\item flat space effects are dominant in the region where the Euclidean distance between any value of the spins is small compared to the length scale $k$ (i.e., $L\ll k$).

\end{itemize}

{\bf Acknowledgement:}~SC would like to thank Junior Scientist position at Max Planck Institute for Gravitational Physics,  Potsdam and J. C. Bose Visiting Scientist position at NISER,  Bhubaneswar.  SP acknowledges the J. C. Bose National Fellowship for support of his
research. SC, NG, RND would like to thank NISER Bhubaneswar, IISER Mohali and IIT Bombay respectively for providing fellowships. Last but not the least, we would like to acknowledge our debt to 
the people belonging to the various part of the world for their generous and steady support for research in natural sciences. Finally, we would like to thank the referees and the editors for many useful suggestions which greatly improved the manuscript.
\\ \\
{\bf Important note:} A detailed supplementary material is added as Appendix to clarify the background material related to the present problem. Some additional plots and results, relevant to the study, are also discussed.

\newpage
\begin{center}
	{\bf Appendix}
\end{center}
\appendix
\section{A. Computation of 
		Wightman functions}
\label{appendix:A}
\begin{figure}[hb]
	\centering
	\includegraphics[width=10cm,height=3.9cm]{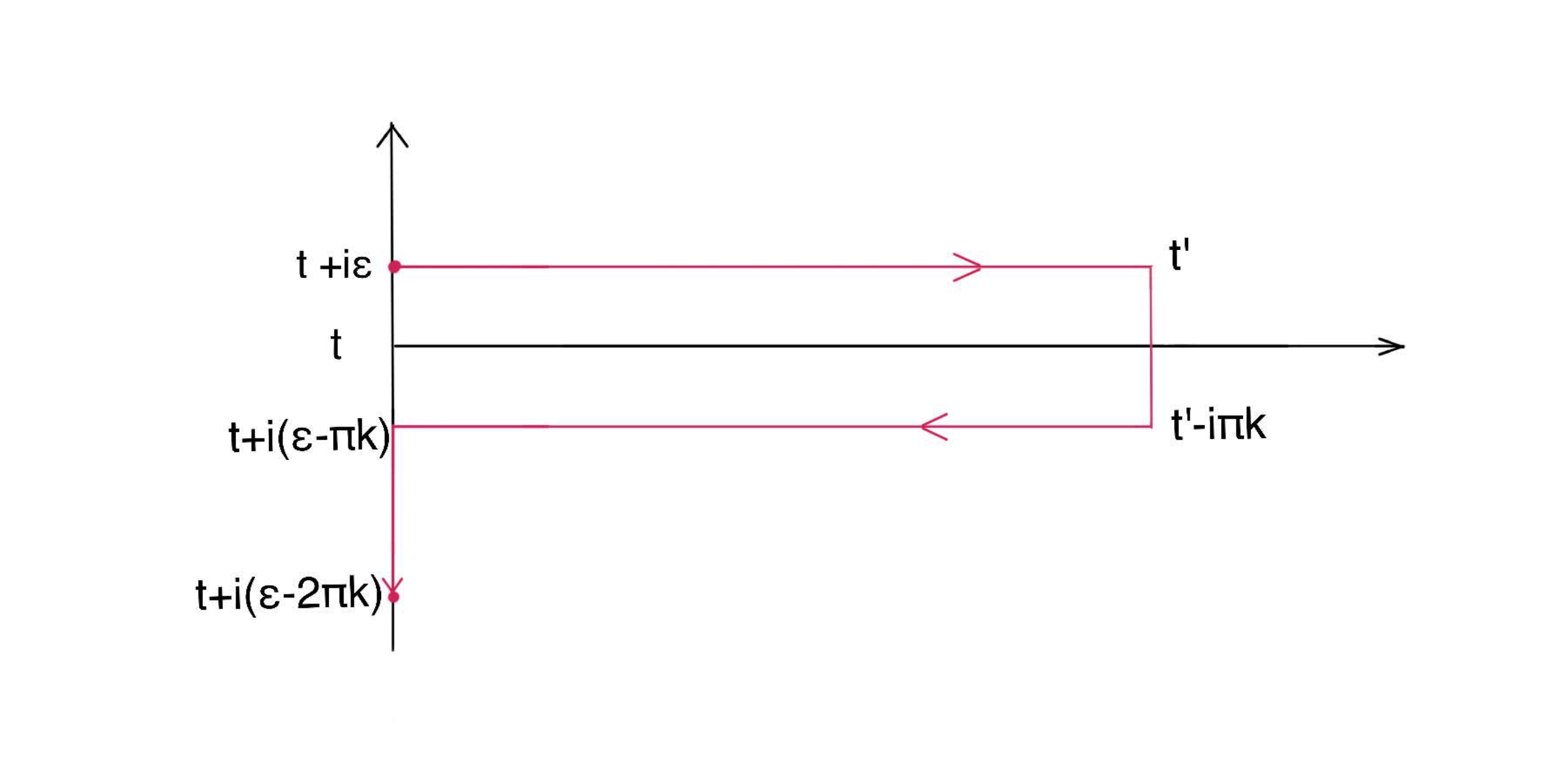}
	\caption{Schwinger Keldysh contour for computing 
		Wightman Functions.}
	\label{fig:my_label}
\end{figure}
To compute the 
Wightman functions of the probe massless scalar field present in the external thermal bath we use the 4D static de Sitter geometry of our space-time as mentioned earlier. In this coordinate system, the equation of motion of the massless external probe scalar field can be written as:
\begin{eqnarray}
	&&\left[\frac{1}{\cosh^3\left(\frac{t}{\alpha}\right)}\partial_{t}\left(\cosh^3\left(\frac{t}{\alpha}\right)\partial_{t}\right)\right.\nonumber\\ && \left.~~~~~~~-\frac{1}{\alpha^2\cosh^2\left(\frac{t}{\alpha}\right)}{\bf L}^2\right]\Phi(t,\chi,\theta,\phi)=0~,
\end{eqnarray}
where ${\bf L}^2$ is the {\it Laplacian operator}, which is defined as:
\begin{eqnarray}
	{\bf L}^2&=&\frac{1}{\sin^2\chi}\left[\frac{\partial}{\partial\chi}\left(\sin^2\chi\frac{\partial}{\partial\chi}\right)\right.\nonumber\\
	&&\left.+\frac{1}{\sin\theta}\frac{\partial}{\partial\theta}\left(\sin\theta\frac{\partial}{\partial\theta}\right)+\frac{1}{\sin^2\theta}\frac{\partial^2}{\partial\phi^2}\right]~,
\end{eqnarray}
where $\chi$ is related to the radial coordinate $r$ as, $ r=\sin\chi.$ 

Further, the complete solution for the massless scalar field is given by:
\begin{widetext}
	\begin{eqnarray}
		\Phi(t,r,\theta,\phi)&=&\sum^{\infty}_{l=0}\sum^{+l}_{m=-l}\Phi_{lm}(t,r,\theta,\phi)=\int^{\infty}_{-\infty}\frac{d\omega}{2\pi}\frac{1}{2\alpha\sqrt{\pi\omega}}\sum^{\infty}_{l=0}\sum^{+l}_{m=-l}\frac{Y_{lm}(\theta,\phi)~e^{-i\omega t}}{\left|\frac{\Gamma\left(l+\frac{3}{2}\right)\Gamma(i\alpha\omega)}{\Gamma\left(\frac{l+3+i\alpha\omega}{2}\right)\Gamma\left(\frac{l+i\alpha\omega}{2}\right)}\right|}\nonumber\\
		&&~~~~~~~~~~~~~~~~~~~~~~~~~~~~~~~~~~~~~~~~~~~~~~~~~~~~\left\{\frac{\Gamma\left(l+\frac{3}{2}\right)\Gamma(i\alpha\omega)}{\Gamma\left(\frac{l+3+i\alpha\omega}{2}\right)\Gamma\left(\frac{l+i\alpha\omega}{2}\right)}\left(1+\frac{r^2}{\alpha^2}\right)^{\frac{i\alpha\omega}{2}}\right.\nonumber\\
		&&~~~~~~~~~~~~~~~~~~~~~~~~~~~~~~~~~~~~~~~~~~~~~~~~~~~~~~~~~~~~\left.~+\frac{\Gamma^{*}\left(l+\frac{3}{2}\right)\Gamma^{*}(i\alpha\omega)}{\Gamma^{*}\left(\frac{l+3+i\alpha\omega}{2}\right)\Gamma^{*}\left(\frac{l+i\alpha\omega}{2}\right)}\left(1+\frac{r^2}{\alpha^2}\right)^{-\frac{i\alpha\omega}{2}}\right\}.\nonumber\\
		&&
	\end{eqnarray}
\end{widetext}
Next, using this classical solution of the field equation the quantum field by the following equation:
\begin{widetext}
	\begin{eqnarray}
		\hat{\Phi}(t,r,\theta,\phi)&=&\sum^{\infty}_{l=0}\sum^{+l}_{m=-l}\left[a_{lm}\Phi_{lm}(t,r,\theta,\phi)+a^{\dagger}_{lm}\Phi^{*}_{lm}(t,r,\theta,\phi)\right].~~~
	\end{eqnarray}
	
\end{widetext}
where the quantum states are defined through the following condition,
\begin{equation}
	a_{lm}|\Psi \rangle=0,~~~~~{\rm where}~~l=0,\cdots,\infty;~~m=-l,\cdots,+l.
\end{equation}
Here $a_{lm}$ and $a^{\dagger}_{lm}$ represent the annihilation and creation operator of the quantum thermal vacuum state $|\Psi\rangle$ which is defined in the bath.

Now, we define the consecutive distance between any two identical static spins localized at the coordinates $(r,\theta,\phi)$ and $(r,\theta^{'},\phi)$ as:
\begin{eqnarray}
	\Delta z^{2}&=&\sum^{4}_{i=1}(z_{i}-z'_{i})^{2}\nonumber\\
	&=&\left(\alpha^2-r^2\right)\left[\cosh\left(\frac{t}{\alpha}\right)-\cosh\left(\frac{t^{'}}{\alpha}\right)\right]^2+L^2,~~~~~
\end{eqnarray}
Here  $L$ represents the euclidean distance between the any two identical spins which is defined as, 
\begin{eqnarray}
	L=2r\sin\left(\frac{\Delta\theta}{2}\right),
\end{eqnarray}
where, $\Delta\theta$ is defined as,
$\Delta\theta=\theta-\theta^{'}.$ 

Further, the 
Wightman function for massless probe scalar field can be expressed as:
\begin{widetext}
	\begin{eqnarray}
		G_{N}(x,x^{'})&=&\begin{pmatrix} ~
			\underbrace{G^{\delta\delta}(x,x')}_{\textcolor{red}{Auto-Correlation}}~~~ &~~~ \underbrace{G^{\delta\eta}(x,x')}_{\textcolor{red}{Cross-Correlation}}~ \\
			~\underbrace{G^{\eta\delta}(x,x')}_{\textcolor{red}{Cross-Correlation}}~~~ &~~~ \underbrace{G^{\eta\eta}(x,x') }_{\textcolor{red}{Auto-Correlation}}~
		\end{pmatrix}_{\beta}=\begin{pmatrix} ~
			\langle \hat{\Phi}({\bf x_{\delta}},\tau)\Phi({\bf x_{\delta}},\tau')\rangle_{\beta}~~~ &~~~ \langle \hat{\Phi}({\bf x_{\delta}},\tau)\Phi({\bf x_{\eta}},\tau')\rangle_{\beta}~ \\ \\
			~\langle \hat{\Phi}({\bf x_{\eta}},\tau)\Phi({\bf x_{\delta}},\tau')\rangle_{\beta}~~~ &~~~ \langle \hat{\Phi}({\bf x_{\eta}},\tau)\Phi({\bf x_{\eta}},\tau')\rangle_{\beta} ~
		\end{pmatrix}~,\nonumber\\
		&& ~~~~~~~~~~~~~~~~~~~~~~~~~~~~~~~~~~~~~~~~~~~~~~~~\forall~~\delta,\eta=1,\cdots,N~{\rm (for~both~even~\&~ odd)}.
	\end{eqnarray} 
\end{widetext}
where the individual Wightman functions can be computed using the well known {\it Schwinger Keldysh} path integral technique as:
\begin{widetext}
	\begin{eqnarray}
		G^{\delta\delta}(x,x')  = G^{\eta\eta}(x,x')
		={\rm Tr}\left[\rho_{B}~ \hat{\Phi}({\bf x_{\delta}},\tau)\hat{\Phi}({\bf x_{\delta}},\tau')\right]
		&=&\langle \Psi|\rho_{B}~ \hat{\Phi}({\bf x_{\delta}},\tau)\hat{\Phi}({\bf x_{\delta}},\tau')|\Psi\rangle \nonumber\\
		&=&- \frac{1}{4\pi^{2}}\frac{1}{\left\{(z_{0}-z'_{0})^{2}-(z_{1}-z'_{1})^{2}-i\epsilon\right\}}\nonumber\\   
		& =&- \frac{1}{16\pi^{2}k^{2}}\frac{1}{\displaystyle \sinh^2\left(\frac{\Delta \tau}{2k}-i\epsilon\right)}, 
		\\
		G^{\delta\eta}(x,x')  = G^{\eta\delta}(x,x')
		={\rm Tr}\left[\rho_{B}~ \hat{\Phi}({\bf x_{\eta}},\tau)\hat{\Phi}({\bf x_{\delta}},\tau')\right]
		&=&\langle \Psi|\rho_{B}~ \hat{\Phi}({\bf x_{\eta}},\tau)\hat{\Phi}({\bf x_{\delta}},\tau')|\Psi\rangle \nonumber\\
		&=&-\frac{1}{4\pi^{2}}\frac{1}{(z_{0}-z'_{0})^{2}-\Delta z^{2}-i\epsilon}\nonumber \\
		& =&-\frac{1}{16\pi^{2}k^{2}}\frac{1}{\displaystyle \left\{\sinh[2](\frac{\Delta \tau}{2k}-i\epsilon)-\frac{r^{2}}{k^{2}}\sin[2](\frac{\Delta \theta}{2})\right\}},~~~~~~
	\end{eqnarray}
\end{widetext}
where we use the result, $\displaystyle \sinh(\frac{\Delta \tau}{2k}-i\epsilon)\sim\sinh(\frac{\Delta \tau}{2k})-i\epsilon\cosh(\frac{\Delta \tau}{2k})$.
Here the thermal density matrix at the bath is defined as:
\begin{equation}
	\rho_{B}=\exp\left(-\beta H_B\right)/Z_B
\end{equation}
where $H_B$ is the bath Hamiltonian of the massless scalar field which is defined in Eqn.~(\ref{wqw}) and $Z_B$ is the partition function of the massless scalar field placed at the thermal bath, defined as:
\begin{eqnarray}
	Z_B={\rm Tr}\left[\exp\left(-\beta H_B\right)\right]=\langle \Psi |\exp\left(-\beta H_B\right)|\Psi\rangle.
\end{eqnarray}
Here $|\Psi\rangle$ is the Bunch Davies thermal state of the bath which is used to compute the trace operation to determine the individual entries of the Wightman functions using {\it Schwinger-Keldysh} technique. However, this result can be generalised to any non  Bunch Davies state (for example, $\alpha$ vacua). Additionally, we define the following quantities:
\begin{eqnarray}
	&&k=\sqrt{g_{00}}\alpha  = \sqrt{\alpha^{2}-r^{2}},\\ &&\Delta \tau=\sqrt{g_{00}}(t-t')=k\left(\frac{t-t'}{\alpha}\right),~~~~~~~~~~
\end{eqnarray}
where $\tau$ is the proper-time and the length scale \bea k=\sqrt{12/R}\eea represents the inverse of curvature in de Sitter static patch.

\section{B. Computation of Hilbert transformation of 
		Wightman functions}
\label{appendix:B}
Now, using the Hilbert transformations one can easily fix the elements of the effective Hamiltonian matrix $H^{(\delta\eta)}_{ij}$ as appearing in the {\it Lamb Shift} part of the Hamiltonian:
\begin{widetext}
	\begin{eqnarray}
		H^{(\delta\eta)}_{ij}=H^{(\eta\delta)}_{ij}=\large
		\left\{
		\begin{array}{lr}
			\displaystyle \mathcal{D}^{\delta\delta}_{1}\delta_{ij}-i\mathcal{Q}^{\delta\delta}_{1}\epsilon_{ijk}\delta_{3k}-\mathcal{D}^{\delta\delta}_{1}\delta_{3i}\delta_{3j}, & \text{$\delta=\eta$}\\ \\ \\
			\displaystyle  \mathcal{D}^{\delta\eta}_{2}\delta_{ij}-i\mathcal{Q}^{\delta\eta}_{2}\epsilon_{ijk}\delta_{3k}-\mathcal{D}^{\delta\eta}_{2}\delta_{3i}\delta_{3j}. & \text{$\delta \neq \eta$}
		\end{array}
		\right.~~~~~
	\end{eqnarray}
\end{widetext}

where we define:
\begin{eqnarray}
	\label{d1z}\mathcal{D}^{\delta\delta}_{1}&=&\frac{\mu^2}{4}\left[{\cal K}^{(\delta\delta)}(\omega_0)+{\cal K}^{(\delta\delta)}(-\omega_0)\right]~~\\
	\label{d2z}\mathcal{Q}^{\delta\delta}_{1}&=&\frac{\mu^2}{4}\left[{\cal K}^{(\delta\delta)}(\omega_0)-{\cal K}^{(\delta\delta)}(-\omega_0)\right],~~\\
	\label{d3z}\mathcal{D}^{\delta\eta}_{2}&=&\frac{\mu^2}{4}\left[{\cal K}^{(\delta\eta)}(\omega_0)+{\cal K}^{(\delta\eta)}(-\omega_0)\right],~~ \\
	\label{d4z}\mathcal{Q}^{\delta\eta}_{2}&=&\frac{\mu^2}{4}\left[{\cal K}^{(\delta\eta)}(\omega_0)-{\cal K}^{(\delta\eta)}(-\omega_0)\right],~~
\end{eqnarray}
where ${\cal K}^{\delta\eta}(\pm \omega_0)\forall (\delta,\eta=1,\cdots,N)$ represents the Hilbert transform of the Wightman functions which can be computed as:
\begin{eqnarray}
	\label{q1z}\mathcal{K}^{\delta\delta}(\pm\omega_{0}) & =& \frac{P}{2\pi^2 i}\int_{-\infty}^{\infty}d\omega~\frac{1}{\omega \mp \omega_{0}}\frac{\omega}{1-e^{2\pi k\omega}},\\
	\label{q2z}\mathcal{K}^{\delta\eta}(\pm\omega_{0}) & =&\frac{P}{2\pi^2 i}\int_{-\infty}^{\infty}d\omega~\frac{{\cal T}(\omega,L/2)}{\omega \mp \omega_{0}}\frac{\omega }{1-e^{2\pi k\omega}}.
\end{eqnarray}
Here, $P$ represents the principal part of the each integrals. For simplicity we also define frequency and euclidean distance dependent a new function ${\cal T}(\omega,L/2)$ as:
\begin{eqnarray}
	{\cal T}(\omega,L/2)=\frac{\sin(2k\omega \sinh^{-1}\left(L/2k\right))}{L\omega\sqrt{1+\left(L/2k\right)^2}}~.
\end{eqnarray}
Finally, substituting the these above mentioned expressions and using {\it Bethe regularisation} technique we get the following simplified results:
\begin{widetext}
	\begingroup
	\begin{eqnarray}
		H^{(\delta\eta)}_{ij}=H^{(\eta\delta)}_{ij}=\frac{\mu^{2}P}{4\pi^2 i}\times
		\large\left\{
		\begin{array}{lr}
			\displaystyle \int_{-\infty}^{\infty}d\omega~\frac{\omega\left\{\left(\delta_{ij}-\delta_{3i}\delta_{3j}\right)\omega-i\epsilon_{ijk}\delta_{3k}\omega_0\right\}}{\left(1-e^{-2\pi k\omega}\right)\left(\omega+\omega_0\right)\left(\omega-\omega_0\right)}=0, & \text{$\delta=\eta$}\\ \\ \\
			\displaystyle \int_{-\infty}^{\infty}d\omega~\frac{\omega\left\{\left(\delta_{ij}-\delta_{3i}\delta_{3j}\right)\omega-i\epsilon_{ijk}\delta_{3k}\omega_0\right\}{\cal T}(\omega,L/2)}{\left(1-e^{-2\pi k\omega}\right)\left(\omega+\omega_0\right)\left(\omega-\omega_0\right)}\\
			\displaystyle=\frac{2\pi}{L\sqrt{1+\left(L/2k\right)^2}}\cos(2k\omega_0 \sinh^{-1}\left(L/2k\right))\\ \displaystyle=\frac{16\pi^2}{\mu^2}{\cal F}(L,k,\omega_0). ~~~~~~~~& \text{$\delta \neq \eta$}
		\end{array}
		\right.~~~~~
	\end{eqnarray}
	\endgroup
\end{widetext}
where the function ${\cal F}(L,k,\omega_0)$ is defined in Eqn.~(\ref{sdsd}).
Hence these matrix elements are fixed which will be needed for the further computation of the spectroscopic shifts from different possible entangled states for the $N$ spin system under consideration. 

{\section{\bf C. States for $N=2$ (even) and $N=3$ (odd) spins}
	\label{Appendix:C}}
For $N=2$ case the sets of eigenstates ($|g_1\rangle,|e_1\rangle$) and ($|g_2\rangle,|e_2\rangle$) are described by the following expressions: 
\begin{eqnarray} && \underline{\textcolor{blue}{\bf For~spin~1:}}\nonumber\\
	&&~~~~~~~H_1=\frac{\omega}{2}\left(\sigma^{1}_{1}\cos\alpha^{1}+\sigma^{1}_{2}\cos\beta^{1}+\sigma^{1}_{3}\cos\gamma^{1}\right)\nonumber\\
	&&\underline{\textcolor{blue}{\bf Ground~state}}\Rightarrow \nonumber\\
	&&|g_1\rangle =N_1\begin{pmatrix} 
		\displaystyle -\frac{\left(\cos\alpha^{1}-i\cos\beta^{1}\right)}{1+\cos\gamma^{1}} \\
		1
	\end{pmatrix}\nonumber\\
	&&\Rightarrow\textcolor{blue}{\bf Eigenvalue}~~E^{(2)}_G=-\frac{\omega}{2},\nonumber\eea
	\bea
	&&\underline{\textcolor{blue}{\bf Excited~state}}\Rightarrow\nonumber\\
	&& |e_1\rangle =N_1\begin{pmatrix} 
		\displaystyle 1\\
		\displaystyle  \frac{\left(\cos\alpha^{1}+i\cos\beta^{1}\right)}{1+\cos\gamma^{1}} ~~
	\end{pmatrix}\nonumber\\
	&&\Rightarrow\textcolor{blue}{\bf Eigenvalue}~~E^{(2)}_E=\frac{\omega}{2}.~~~~~~~~~~~~ \\
	\nonumber\\
	&&  \underline{\textcolor{blue}{\bf For~spin~2:}}\nonumber\\
	&&~~~~~~~H_1=\frac{\omega}{2}\left(\sigma^{2}_{1}\cos\alpha^{2}+\sigma^{2}_{2}\cos\beta^{2}+\sigma^{2}_{3}\cos\gamma^{2}\right)\nonumber\\
	&&\underline{\textcolor{blue}{\bf Ground~state}}\Rightarrow \nonumber\\
	&&|g_2\rangle =N_2\begin{pmatrix} 
		\displaystyle -\frac{\left(\cos\alpha^{2}-i\cos\beta^{2}\right)}{1+\cos\gamma^{2}} \\
		1
	\end{pmatrix}\nonumber\\
	&&\Rightarrow\textcolor{blue}{\bf Eigenvalue}~~E^{(2)}_G=-\frac{\omega}{2},\\
	&&\underline{\textcolor{blue}{\bf Excited~state}}\Rightarrow\nonumber\\
	&& |e_2\rangle =N_2\large\begin{pmatrix} 
		1\\
		\displaystyle \frac{\left(\cos\alpha^{2}+i\cos\beta^{2}\right)}{1+\cos\gamma^{2}} ~~
	\end{pmatrix}\nonumber\\
	&&\Rightarrow\textcolor{blue}{\bf Eigenvalue}~~E^{(2)}_E=\frac{\omega}{2},~~~~~~~~~~~~
\end{eqnarray}
where we define the normalisation factor for spin $1$ and $2$ as:
\begin{eqnarray}
	&& N_1=\frac{1}{\sqrt{2}}\sqrt{1+\cos\gamma^{1}},\\
	&& N_2=\frac{1}{\sqrt{2}}\sqrt{1+\cos\gamma^{2}}.
\end{eqnarray}
Consequently, the ground ($|{G}\rangle$), excited ($|{E}\rangle$), symmetric ($|{S}\rangle$) and the anti-symmetric ($|{A}\rangle$) state of the two-entangled spin system can be expressed by the following expression:
\begin{eqnarray}
	&&\underline{\textcolor{blue}{\bf Ground ~state:}}\Rightarrow\nonumber\\
	&&|G \rangle = |g_1 \rangle \otimes| g_2 \rangle\nonumber\\
	&& ={\cal N}_{1,2}\large\begin{pmatrix} 
		\displaystyle -\frac{\left(\cos\alpha^{1}-i\cos\beta^{1}\right)}{1+\cos\gamma^{1}}\frac{\left(\cos\alpha^{2}-i\cos\beta^{2}\right)}{1+\cos\gamma^{2}} \\
		\displaystyle -\frac{\left(\cos\alpha^{1}-i\cos\beta^{1}\right)}{1+\cos\gamma^{1}}\\
		\displaystyle -\frac{\left(\cos\alpha^{2}-i\cos\beta^{2}\right)}{1+\cos\gamma^{2}} \\
		1 
	\end{pmatrix},~~~\\
	&&\underline{\textcolor{blue}{\bf Excited ~state:}}\Rightarrow\nonumber\\
	&&|E \rangle = |e_1 \rangle \otimes |e_2 \rangle\nonumber\\
	&&   ={\cal N}_{1,2}\large\begin{pmatrix} 
		\displaystyle 1 \\
		\displaystyle \frac{\left(\cos\alpha^{2}+i\cos\beta^{2}\right)}{1+\cos\gamma^{2}}\\
		\displaystyle \frac{\left(\cos\alpha^{1}+i\cos\beta^{1}\right)}{1+\cos\gamma^{1}} \\
		\displaystyle \frac{\left(\cos\alpha^{1}+i\cos\beta^{1}\right)}{1+\cos\gamma^{1}}\frac{\left(\cos\alpha^{2}+i\cos\beta^{2}\right)}{1+\cos\gamma^{2}}
	\end{pmatrix},~~~~~~~~, \eea
	\bea &&\underline{\textcolor{blue}{\bf Symmetric ~state:}}\Rightarrow\nonumber\\
	&&|S \rangle = \frac{1}{\sqrt{2}} [ |e_1 \rangle \otimes |g_2 \rangle + |g_1 \rangle \otimes |e_2 \rangle]  \nonumber\\
	&&= \frac{{\cal N}_{1,2}}{\sqrt{2}}\large\begin{pmatrix} 
		\displaystyle -\frac{\left(\cos\alpha^{1}-i\cos\beta^{1}\right)}{1+\cos\gamma^{1}}-\frac{\left(\cos\alpha^{2}-i\cos\beta^{2}\right)}{1+\cos\gamma^{2}} \\
		\displaystyle 1-\frac{\left(\cos\alpha^{1}-i\cos\beta^{1}\right)}{1+\cos\gamma^{1}}\frac{\left(\cos\alpha^{2}+i\cos\beta^{2}\right)}{1+\cos\gamma^{2}} \\
		\displaystyle 1-\frac{\left(\cos\alpha^{1}+i\cos\beta^{1}\right)}{1+\cos\gamma^{1}}\frac{\left(\cos\alpha^{2}-i\cos\beta^{2}\right)}{1+\cos\gamma^{2}}  \\
		\displaystyle \frac{\left(\cos\alpha^{1}+i\cos\beta^{1}\right)}{1+\cos\gamma^{1}}+\frac{\left(\cos\alpha^{2}+i\cos\beta^{2}\right)}{1+\cos\gamma^{2}} 
	\end{pmatrix},~~~~~, \\
	&&\underline{\textcolor{blue}{\bf Antisymmetric ~state:}}\Rightarrow\nonumber\\
	&&|A \rangle = \frac{1}{\sqrt{2}} [ |e_1 \rangle \otimes |g_2 \rangle - |g_1 \rangle \otimes |e_2 \rangle]  \nonumber\\
	&&= \frac{{\cal N}_{1,2}}{\sqrt{2}}\large\begin{pmatrix} 
		\displaystyle \frac{\left(\cos\alpha^{1}-i\cos\beta^{1}\right)}{1+\cos\gamma^{1}}-\frac{\left(\cos\alpha^{2}-i\cos\beta^{2}\right)}{1+\cos\gamma^{2}} \\
		\displaystyle 1+\frac{\left(\cos\alpha^{1}-i\cos\beta^{1}\right)}{1+\cos\gamma^{1}}\frac{\left(\cos\alpha^{2}+i\cos\beta^{2}\right)}{1+\cos\gamma^{2}} \\
		\displaystyle -1-\frac{\left(\cos\alpha^{1}+i\cos\beta^{1}\right)}{1+\cos\gamma^{1}}\frac{\left(\cos\alpha^{2}-i\cos\beta^{2}\right)}{1+\cos\gamma^{2}}  \\
		\displaystyle \frac{\left(\cos\alpha^{1}+i\cos\beta^{1}\right)}{1+\cos\gamma^{1}}-\frac{\left(\cos\alpha^{2}+i\cos\beta^{2}\right)}{1+\cos\gamma^{2}} 
	\end{pmatrix},~~~~, 
\end{eqnarray}
where we define the two spin normalisation factor ${\cal N}_{1,2}$ as:
\begin{eqnarray}
	{\cal N}_{1,2}=N_1N_2=\frac{1}{2}\sqrt{(1+\cos\gamma^{1})(1+\cos\gamma^{2})}.
\end{eqnarray}
For $N=3$ case for the third spin
the sets of eigenstates ($|g_3\rangle,|e_3\rangle$) are described by the following expressions: 
sets of eigenstates ($|g_1\rangle,|e_1\rangle$) and ($|g_2\rangle,|e_2\rangle$) are described by the following expressions: 
\begin{eqnarray} && \underline{\textcolor{blue}{\bf For~spin~1:}}\nonumber\\
	&&~~~~~~~H_1=\frac{\omega}{2}\left(\sigma^{1}_{1}\cos\alpha^{1}+\sigma^{1}_{2}\cos\beta^{1}+\sigma^{1}_{3}\cos\gamma^{1}\right)\nonumber\\
	&&\underline{\textcolor{blue}{\bf Ground~state}}\Rightarrow \nonumber\\
	&&|g_1\rangle =N_1\begin{pmatrix} 
		\displaystyle -\frac{\left(\cos\alpha^{1}-i\cos\beta^{1}\right)}{1+\cos\gamma^{1}} \\
		1
	\end{pmatrix}\nonumber\\
	&&\Rightarrow\textcolor{blue}{\bf Eigenvalue}~~E^{(2)}_G=-\frac{\omega}{2},\\
	&&\underline{\textcolor{blue}{\bf Excited~state}}\Rightarrow\nonumber\\
	&& |e_1\rangle =N_1\begin{pmatrix} 
		\displaystyle 1\\
		\displaystyle  \frac{\left(\cos\alpha^{1}+i\cos\beta^{1}\right)}{1+\cos\gamma^{1}} ~~
	\end{pmatrix}\nonumber\\
	&&\Rightarrow\textcolor{blue}{\bf Eigenvalue}~~E^{(2)}_E=\frac{\omega}{2}.~~~~~~~~~~~~ \\
	\nonumber\eea
	\bea
	&&  \underline{\textcolor{blue}{\bf For~spin~2:}}\nonumber\\
	&&~~~~~~~H_1=\frac{\omega}{2}\left(\sigma^{2}_{1}\cos\alpha^{2}+\sigma^{2}_{2}\cos\beta^{2}+\sigma^{2}_{3}\cos\gamma^{2}\right)\nonumber\\
	&&\underline{\textcolor{blue}{\bf Ground~state}}\Rightarrow \nonumber\\
	&&|g_2\rangle =N_2\large\displaystyle\begin{pmatrix} 
		\displaystyle -\frac{\left(\cos\alpha^{2}-i\cos\beta^{2}\right)}{1+\cos\gamma^{2}} \\
		1
	\end{pmatrix}\nonumber\\
	&&\Rightarrow\textcolor{blue}{\bf Eigenvalue}~~E^{(2)}_G=-\frac{\omega}{2},\\
	&&\underline{\textcolor{blue}{\bf Excited~state}}\Rightarrow\nonumber\\
	&& |e_2\rangle =N_2\large\begin{pmatrix} 
		1\\
		\displaystyle \frac{\left(\cos\alpha^{2}+i\cos\beta^{2}\right)}{1+\cos\gamma^{2}} ~~
	\end{pmatrix}\nonumber\\
	&&\Rightarrow\textcolor{blue}{\bf Eigenvalue}~~E^{(2)}_E=\frac{\omega}{2},~~~~~~~~~~~~
\end{eqnarray}

\begin{eqnarray} && \underline{\textcolor{blue}{\bf For~spin~3:}}\nonumber\\
	&&~~~~~~~H_3=\frac{\omega}{2}\left(\sigma^{3}_{1}\cos\alpha^{3}+\sigma^{3}_{2}\cos\beta^{3}+\sigma^{3}_{3}\cos\gamma^{3}\right)\nonumber\\
	&&\underline{\textcolor{blue}{\bf Ground~state}}\Rightarrow \nonumber\\
	&&|g_3\rangle =N_3\large\begin{pmatrix} 
		\displaystyle -\frac{\left(\cos\alpha^{3}-i\cos\beta^{3}\right)}{1+\cos\gamma^{3}} \\
		1
	\end{pmatrix}\nonumber\\
	&&\Rightarrow\textcolor{blue}{\bf Eigenvalue}~~E^{(3)}_G=-\frac{\omega}{2},\eea
	\bea
	&&\underline{\textcolor{blue}{\bf Excited~state}}\Rightarrow\nonumber\\
	&& |e_3\rangle =N_3\large\begin{pmatrix} 
		\displaystyle 1\\
		\displaystyle  \frac{\left(\cos\alpha^{3}+i\cos\beta^{3}\right)}{1+\cos\gamma^{3}} ~~
	\end{pmatrix}\nonumber\\
	&&\Rightarrow\textcolor{blue}{\bf Eigenvalue}~~E^{(3)}_E=\frac{\omega}{2}.~~~~~~~~~~~~ \\
	\nonumber\end{eqnarray}
where we define the normalisation factor for spin $1$, $2$ and $3$ as:
\begin{eqnarray}
	&& N_{\delta}=\frac{1}{\sqrt{2}}\sqrt{1+\cos\gamma^{1}}.
\end{eqnarray}
Consequently, the ground  ($|{G}\rangle$), excited ($|{E}\rangle$), symmetric ($|{S}\rangle$) and the anti-symmetric ($|{A}\rangle$) state of the three-entangled spin system  can be expressed as:
\begin{widetext}
	\begin{eqnarray}
		&&\underline{\textcolor{blue}{\bf Ground ~state:}}\Rightarrow\nonumber\\
		&&|G \rangle = \frac{1}{\sqrt{3}} [|g_1 \rangle \otimes |g_2 \rangle +|g_1 \rangle \otimes |g_3 \rangle + |g_2 \rangle \otimes |g_3 \rangle]  =\frac{1}{2\sqrt{3}}\large\left(
		\begin{array}{c}
			\scriptscriptstyle \frac{(\cos (\text{$\alpha $1})-i \cos (\text{$\beta $1})) (\cos (\text{$\alpha $2})-i \cos (\text{$\beta $2}))}{ \sqrt{\cos (\text{$\gamma $1})+1} \sqrt{\cos (\text{$\gamma $2})+1}}+\frac{(\cos (\text{$\alpha $1})-i \cos (\text{$\beta $1})) (\cos (\text{$\alpha $3})-i \cos (\text{$\beta $3}))}{ \sqrt{\cos (\text{$\gamma $1})+1} \sqrt{\cos (\text{$\gamma $3})+1}}\\ \scriptscriptstyle+\frac{(\cos (\text{$\alpha $2})-i \cos (\text{$\beta $2})) (\cos (\text{$\alpha $3})-i \cos (\text{$\beta $3}))}{ \sqrt{\cos (\text{$\gamma $2})+1} \sqrt{\cos (\text{$\gamma $3})+1}} \\ 
			\scriptscriptstyle -\frac{\sqrt{\cos (\text{$\gamma $2})+1} (\cos (\text{$\alpha $1})-i \cos (\text{$\beta $1}))}{ \sqrt{\cos (\text{$\gamma $1})+1}}-\frac{\sqrt{\cos (\text{$\gamma $3})+1} (\cos (\text{$\alpha $2})-i \cos (\text{$\beta $2}))}{ \sqrt{\cos (\text{$\gamma $2})+1}}\\\scriptscriptstyle-\frac{\sqrt{\cos (\text{$\gamma $1})+1} (\cos (\text{$\alpha $3})-i \cos (\text{$\beta $3}))}{ \sqrt{\cos (\text{$\gamma $3})+1}}\\ 
			\scriptscriptstyle -\frac{\sqrt{\cos (\text{$\gamma $3})+1} (\cos (\text{$\alpha $1})-i \cos (\text{$\beta $1}))}{ \sqrt{\cos (\text{$\gamma $1})+1}}-\frac{\sqrt{\cos (\text{$\gamma $1})+1} (\cos (\text{$\alpha $2})-i \cos (\text{$\beta $2}))}{ \sqrt{\cos (\text{$\gamma $2})+1}}\\\scriptscriptstyle-\frac{\sqrt{\cos (\text{$\gamma $2})+1} (\cos (\text{$\alpha $3})-i \cos (\text{$\beta $3}))}{ \sqrt{\cos (\text{$\gamma $3})+1}} \\ 
			\scriptscriptstyle  \sqrt{\cos (\text{$\gamma $1})+1} \sqrt{\cos (\text{$\gamma $2})+1}+\sqrt{\cos (\text{$\gamma $1})+1} \sqrt{\cos (\text{$\gamma $3})+1}+ \sqrt{\cos (\text{$\gamma $2})+1} \sqrt{\cos (\text{$\gamma $3})+1} \\
		\end{array}
		\right),~~~~~\end{eqnarray}
	\begin{eqnarray}
		&&\underline{\textcolor{blue}{\bf Excited ~state:}}\Rightarrow\nonumber\\
		&&|E \rangle =\frac{1}{\sqrt{3}} [|e_1 \rangle \otimes |e_2 \rangle +|e_1 \rangle \otimes |e_3 \rangle + |e_2 \rangle \otimes |e_3 \rangle]   =\frac{1}{2\sqrt{3}}\large\left(
		\begin{array}{c}
			\scriptscriptstyle\sqrt{\cos (\text{$\gamma $1})+1} \sqrt{\cos (\text{$\gamma $2})+1}+ \sqrt{\cos (\text{$\gamma $1})+1} \sqrt{\cos (\text{$\gamma $3})+1} + \sqrt{\cos (\text{$\gamma $2})+1} \sqrt{\cos (\text{$\gamma $3})+1}\\
			\scriptscriptstyle \frac{\sqrt{\cos (\text{$\gamma $3})+1} (\cos (\text{$\alpha $1})+i \cos (\text{$\beta $1}))}{ \sqrt{\cos (\text{$\gamma $1})+1}}+\frac{\sqrt{\cos (\text{$\gamma $1})+1} (\cos (\text{$\alpha $2})+i \cos (\text{$\beta $2}))}{ \sqrt{\cos (\text{$\gamma $2})+1}}\\ \scriptscriptstyle+\frac{\sqrt{\cos (\text{$\gamma $2})+1} (\cos (\text{$\alpha $3})+i \cos (\text{$\beta $3}))}{ \sqrt{\cos (\text{$\gamma $3})+1}} \\
			\scriptscriptstyle \frac{\sqrt{\cos (\text{$\gamma $2})+1} (\cos (\text{$\alpha $1})+i \cos (\text{$\beta $1}))}{ \sqrt{\cos (\text{$\gamma $1})+1}}+\frac{\sqrt{\cos (\text{$\gamma $3})+1} (\cos (\text{$\alpha $2})+i \cos (\text{$\beta $2}))}{\sqrt{\cos (\text{$\gamma $2})+1}}\\ \scriptscriptstyle+\frac{\sqrt{\cos (\text{$\gamma $1})+1} (\cos (\text{$\alpha $3})+i \cos (\text{$\beta $3}))}{2 \sqrt{\cos (\text{$\gamma $3})+1}}\\
			\scriptscriptstyle \frac{(\cos (\text{$\alpha $1})+i \cos (\text{$\beta $1})) (\cos (\text{$\alpha $2})+i \cos (\text{$\beta $2}))}{\sqrt{\cos (\text{$\gamma $1})+1} \sqrt{\cos (\text{$\gamma $2})+1}}+\frac{(\cos (\text{$\alpha $1})+i \cos (\text{$\beta $1})) (\cos (\text{$\alpha $3})+i \cos (\text{$\beta $3}))}{ \sqrt{\cos (\text{$\gamma $1})+1} \sqrt{\cos (\text{$\gamma $3})+1}}\\ \scriptscriptstyle+\frac{(\cos (\text{$\alpha $2})+i \cos (\text{$\beta $2})) (\cos (\text{$\alpha $3})+i \cos (\text{$\beta $3}))}{ \sqrt{\cos (\text{$\gamma $2})+1} \sqrt{\cos (\text{$\gamma $3})+1}}\\
		\end{array}
		\right),~~~~~, \end{eqnarray}
\end{widetext} 

\begin{widetext} 
	\begin{eqnarray} 
		&&\underline{\textcolor{blue}{\bf Symmetric ~state:}}\Rightarrow\nonumber\\
		&&|S \rangle = \frac{1}{\sqrt{6}} [ |e_1 \rangle \otimes |g_2 \rangle + |g_1 \rangle \otimes |e_2 \rangle + |e_1 \rangle \otimes |g_3 \rangle + |g_1 \rangle \otimes |e_3 \rangle + |e_2 \rangle \otimes |g_3 \rangle + |g_2 \rangle \otimes |e_3 \rangle]  \nonumber\\
		&&=\frac{1}{2 \sqrt{6}}\scalebox{0.00002}{0.00002}{\tiny \left(\setlength\arraycolsep{0.1pt}
			\begin{array}{c}
				\scriptscriptstyle -\frac{\sqrt{\cos (\text{$\gamma $2})+1} (\cos (\text{$\alpha $1})-i \cos (\text{$\beta $1}))}{2 \sqrt{\cos (\text{$\gamma $1})+1}}-\frac{\sqrt{\cos (\text{$\gamma $3})+1} (\cos (\text{$\alpha $1})-i \cos (\text{$\beta $1}))}{ \sqrt{\cos (\text{$\gamma $1})+1}}-\frac{\sqrt{\cos (\text{$\gamma $1})+1} (\cos (\text{$\alpha $2})-i \cos (\text{$\beta $2}))}{ \sqrt{\cos (\text{$\gamma $2})+1}}\\-\frac{\sqrt{\cos (\text{$\gamma $3})+1} (\cos (\text{$\alpha $2})-i \cos (\text{$\beta $2}))}{ \sqrt{\cos (\text{$\gamma $2})+1}}-\frac{\sqrt{\cos (\text{$\gamma $1})+1} (\cos (\text{$\alpha $3})-i \cos (\text{$\beta $3}))}{ \sqrt{\cos (\text{$\gamma $3})+1}}-\frac{\sqrt{\cos (\text{$\gamma $2})+1} (\cos (\text{$\alpha $3})-i \cos (\text{$\beta $3}))}{ \sqrt{\cos (\text{$\gamma $3})+1}} \\ \\ \\
				\scriptscriptstyle-\frac{(\cos (\text{$\alpha $1})-i \cos (\text{$\beta $1})) (\cos (\text{$\alpha $2})+i \cos (\text{$\beta $2}))}{ \sqrt{\cos (\text{$\gamma $1})+1} \sqrt{\cos (\text{$\gamma $2})+1}}-\frac{(\cos (\text{$\alpha $1})-i \cos (\text{$\beta $1})) (\cos (\text{$\alpha $3})+i \cos (\text{$\beta $3}))}{ \sqrt{\cos (\text{$\gamma $1})+1} \sqrt{\cos (\text{$\gamma $3})+1}}-\frac{(\cos (\text{$\alpha $2})-i \cos (\text{$\beta $2})) (\cos (\text{$\alpha $3})+i \cos (\text{$\beta $3}))}{ \sqrt{\cos (\text{$\gamma $2})+1} \sqrt{\cos (\text{$\gamma $3})+1}}\\
				+ \sqrt{\cos (\text{$\gamma $1})+1} \sqrt{\cos (\text{$\gamma $2})+1}+\sqrt{\cos (\text{$\gamma $1})+1} \sqrt{\cos (\text{$\gamma $3})+1}+\sqrt{\cos (\text{$\gamma $2})+1} \sqrt{\cos (\text{$\gamma $3})+1} \\ \\ \\
				\scriptscriptstyle -\frac{(\cos (\text{$\alpha $1})+i \cos (\text{$\beta $1})) (\cos (\text{$\alpha $2})-i \cos (\text{$\beta $2}))}{ \sqrt{\cos (\text{$\gamma $1})+1} \sqrt{\cos (\text{$\gamma $2})+1}}-\frac{(\cos (\text{$\alpha $1})+i \cos (\text{$\beta $1})) (\cos (\text{$\alpha $3})-i \cos (\text{$\beta $3}))}{ \sqrt{\cos (\text{$\gamma $1})+1} \sqrt{\cos (\text{$\gamma $3})+1}}-\frac{(\cos (\text{$\alpha $2})+i \cos (\text{$\beta $2})) (\cos (\text{$\alpha $3})-i \cos (\text{$\beta $3}))}{ \sqrt{\cos (\text{$\gamma $2})+1} \sqrt{\cos (\text{$\gamma $3})+1}}\\+ \sqrt{\cos (\text{$\gamma $1})+1} \sqrt{\cos (\text{$\gamma $2})+1}+ \sqrt{\cos (\text{$\gamma $1})+1} \sqrt{\cos (\text{$\gamma $3})+1}+ \sqrt{\cos (\text{$\gamma $2})+1} \sqrt{\cos (\text{$\gamma $3})+1} \\ \\ \\
				\scriptscriptstyle \frac{\sqrt{\cos (\text{$\gamma $2})+1} (\cos (\text{$\alpha $1})+i \cos (\text{$\beta $1}))}{ \sqrt{\cos (\text{$\gamma $1})+1}}+\frac{\sqrt{\cos (\text{$\gamma $3})+1} (\cos (\text{$\alpha $1})+i \cos (\text{$\beta $1}))}{ \sqrt{\cos (\text{$\gamma $1})+1}}+\frac{\sqrt{\cos (\text{$\gamma $1})+1} (\cos (\text{$\alpha $2})+i \cos (\text{$\beta $2}))}{ \sqrt{\cos (\text{$\gamma $2})+1}}\\+\frac{\sqrt{\cos (\text{$\gamma $3})+1} (\cos (\text{$\alpha $2})+i \cos (\text{$\beta $2}))}{ \sqrt{\cos (\text{$\gamma $2})+1}}+\frac{\sqrt{\cos (\text{$\gamma $1})+1} (\cos (\text{$\alpha $3})+i \cos (\text{$\beta $3}))}{ \sqrt{\cos (\text{$\gamma $3})+1}}+\frac{\sqrt{\cos (\text{$\gamma $2})+1} (\cos (\text{$\alpha $3})+i \cos (\text{$\beta $3}))}{ \sqrt{\cos (\text{$\gamma $3})+1}} \\
			\end{array}
			\right)},~~~~~~~~, \\
		&&\underline{\textcolor{blue}{\bf Antisymmetric ~state:}}\Rightarrow\nonumber\\
		&&|A \rangle =  \frac{1}{\sqrt{6}} [ |e_1 \rangle \otimes |g_2 \rangle - |g_1 \rangle \otimes |e_2 \rangle + |e_1 \rangle \otimes |g_3 \rangle - |g_1 \rangle \otimes |e_3 \rangle + |e_2 \rangle \otimes |g_3 \rangle - |g_2 \rangle \otimes |e_3 \rangle]   \nonumber\\
		&&=\frac{1}{2\sqrt{6}}\scalebox{0.00002}{0.00002}{ \tiny\left(
			\begin{array}{c}
				\scriptscriptstyle\frac{\sqrt{\cos (\text{$\gamma $2})+1} (\cos (\text{$\alpha $1})-i \cos (\text{$\beta $1}))}{ \sqrt{\cos (\text{$\gamma $1})+1}}+\frac{\sqrt{\cos (\text{$\gamma $3})+1} (\cos (\text{$\alpha $1})-i \cos (\text{$\beta $1}))}{ \sqrt{\cos (\text{$\gamma $1})+1}}-\frac{\sqrt{\cos (\text{$\gamma $1})+1} (\cos (\text{$\alpha $2})-i \cos (\text{$\beta $2}))}{ \sqrt{\cos (\text{$\gamma $2})+1}}\\+\frac{\sqrt{\cos (\text{$\gamma $3})+1} (\cos (\text{$\alpha $2})-i \cos (\text{$\beta $2}))}{ \sqrt{\cos (\text{$\gamma $2})+1}}-\frac{\sqrt{\cos (\text{$\gamma $1})+1} (\cos (\text{$\alpha $3})-i \cos (\text{$\beta $3}))}{ \sqrt{\cos (\text{$\gamma $3})+1}}-\frac{\sqrt{\cos (\text{$\gamma $2})+1} (\cos (\text{$\alpha $3})-i \cos (\text{$\beta $3}))}{ \sqrt{\cos (\text{$\gamma $3})+1}} \\ \\ \\
				\scriptscriptstyle \frac{(\cos (\text{$\alpha $1})-i \cos (\text{$\beta $1})) (\cos (\text{$\alpha $2})+i \cos (\text{$\beta $2}))}{ \sqrt{\cos (\text{$\gamma $1})+1} \sqrt{\cos (\text{$\gamma $2})+1}}+\frac{(\cos (\text{$\alpha $1})-i \cos (\text{$\beta $1})) (\cos (\text{$\alpha $3})+i \cos (\text{$\beta $3}))}{ \sqrt{\cos (\text{$\gamma $1})+1} \sqrt{\cos (\text{$\gamma $3})+1}}+\frac{(\cos (\text{$\alpha $2})-i \cos (\text{$\beta $2})) (\cos (\text{$\alpha $3})+i \cos (\text{$\beta $3}))}{ \sqrt{\cos (\text{$\gamma $2})+1} \sqrt{\cos (\text{$\gamma $3})+1}} \\+ \sqrt{\cos (\text{$\gamma $1})+1} \sqrt{\cos (\text{$\gamma $2})+1}+ \sqrt{\cos (\text{$\gamma $1})+1} \sqrt{\cos (\text{$\gamma $3})+1}+\sqrt{\cos (\text{$\gamma $2})+1} \sqrt{\cos (\text{$\gamma $3})+1} \\ \\ \\
				\scriptscriptstyle -\frac{(\cos (\text{$\alpha $1})+i \cos (\text{$\beta $1})) (\cos (\text{$\alpha $2})-i \cos (\text{$\beta $2}))}{ \sqrt{\cos (\text{$\gamma $1})+1} \sqrt{\cos (\text{$\gamma $2})+1}}-\frac{(\cos (\text{$\alpha $1})+i \cos (\text{$\beta $1})) (\cos (\text{$\alpha $3})-i \cos (\text{$\beta $3}))}{\sqrt{\cos (\text{$\gamma $1})+1} \sqrt{\cos (\text{$\gamma $3})+1}}-\frac{(\cos (\text{$\alpha $2})+i \cos (\text{$\beta $2})) (\cos (\text{$\alpha $3})-i \cos (\text{$\beta $3}))}{ \sqrt{\cos (\text{$\gamma $2})+1} \sqrt{\cos (\text{$\gamma $3})+1}}\\- \sqrt{\cos (\text{$\gamma $1})+1} \sqrt{\cos (\text{$\gamma $2})+1}- \sqrt{\cos (\text{$\gamma $1})+1} \sqrt{\cos (\text{$\gamma $3})+1}- \sqrt{\cos (\text{$\gamma $2})+1} \sqrt{\cos (\text{$\gamma $3})+1} \\ \\ \\
				\scriptscriptstyle \frac{\sqrt{\cos (\text{$\gamma $2})+1} (\cos (\text{$\alpha $1})+i \cos (\text{$\beta $1}))}{\sqrt{\cos (\text{$\gamma $1})+1}}+\frac{\sqrt{\cos (\text{$\gamma $3})+1} (\cos (\text{$\alpha $1})+i \cos (\text{$\beta $1}))}{ \sqrt{\cos (\text{$\gamma $1})+1}}-\frac{\sqrt{\cos (\text{$\gamma $1})+1} (\cos (\text{$\alpha $2})+i \cos (\text{$\beta $2}))}{ \sqrt{\cos (\text{$\gamma $2})+1}}\\+\frac{\sqrt{\cos (\text{$\gamma $3})+1} (\cos (\text{$\alpha $2})+i \cos (\text{$\beta $2}))}{ \sqrt{\cos (\text{$\gamma $2})+1}}-\frac{\sqrt{\cos (\text{$\gamma $1})+1} (\cos (\text{$\alpha $3})+i \cos (\text{$\beta $3}))}{ \sqrt{\cos (\text{$\gamma $3})+1}}-\frac{\sqrt{\cos (\text{$\gamma $2})+1} (\cos (\text{$\alpha $3})+i \cos (\text{$\beta $3}))}{ \sqrt{\cos (\text{$\gamma $3})+1}} \\
			\end{array}
			\right)},~~~~~~~~, 
	\end{eqnarray}
\end{widetext}
{\section{\bf D. Direction cosine dependent angular distribution factors for $N=2$ (even) and $N=3$ (odd) spins}
	\label{Appendix:D}}
For $N=2$ case we have two angular distribution $\Gamma_{1;{\cal DC}}$ and $\Gamma_{2;{\cal DC}}$, which are defined as:
\begin{widetext}
	\begin{eqnarray} \Gamma_{1;{\cal DC}}&=&   \Omega\left\{ (B^{2}+C^{2}-A^{2}-D^{2}) \cos(\alpha^{1})\cos(\alpha^{2}) +  (A^{2}+B^{2}+C^{2}+D^{2})\cos(\beta^{1})\cos(\beta^{2})\right\},~~~~~~\\
		\Gamma_{2;{\cal DC}}&=& \Omega\left\{ (\tilde  D^{2}+\tilde A^{2}-\tilde B^{2}-\tilde C^{2}) \cos(\alpha^{1})\cos(\alpha^{2}) -  (\tilde  A^{2}+\tilde B^{2}+\tilde C^{2}+\tilde D^{2})\cos(\beta^{1})\cos(\beta^{2}) \right\},\end{eqnarray}
\end{widetext}
where we define few quantities important for rest of the calculation:
\begin{widetext}
	\begin{eqnarray}
		\displaystyle
		&& A=\left[\frac{\cos \ag_{1}-i \cos \bg_{1}}{1+\cos \gamma_{1}}+\frac{\cos \ag_{2}-i \cos\bg_{2}}{1+\cos \gamma_{2}}\right] \\
		\displaystyle
		&& B=\left[1-\frac{\cos \ag_{1}-i \cos \bg_{1}}{1+\cos \gamma_{1}}.\frac{\cos\ag_{2}+i \cos\bg_{2}}{1+\cos \gamma_{2}}\right] \\
		\displaystyle
		&& C=\left[1-\frac{\cos \ag_{1}+i \cos\bg_{1}}{1+\cos \gamma_{1}}.\frac{\cos \ag_{2}-i\cos \bg_{2}}{1+\cos \gamma_{2}}\right] \\
		\displaystyle
		&& D=\left[\frac{\cos \ag_{1}+i \cos \bg_{1}}{1+\cos \gamma_{1}}+\frac{\cos \ag_{2}+i \cos \bg_{2}}{1+\cos \gamma_{2}}\right] \end{eqnarray}
\end{widetext}
\begin{widetext}
	\begin{eqnarray}
		&& \tilde A=\left[\frac{\cos\ag_{1}-i \cos\bg_{1}}{1+\cos \gamma_{1}}-\frac{\cos\ag_{2}-i \cos\bg_{2}}{1+\cos\gamma_{2}}\right],\\
		\displaystyle
		&& \tilde B=\left[1+\frac{\cos\ag_{1}-i \cos\bg_{1}}{1+\cos \gamma_{1}}\frac{\cos\ag_{2}+i \cos\bg_{2}}{1+\cos\gamma_{2}}\right],\\
		\displaystyle
		&& \tilde C=\left[-1-\frac{\cos\ag_{1}+ i \cos\bg_{1}}{1+\cos \gamma_{1}}\frac{\cos\ag_{2}-i \cos\bg_{2}}{1+\cos\gamma_{2}}\right],\\
		\displaystyle
		&& \tilde D=\left[\frac{\cos\ag_{1}-i \cos\bg_{1}}{1+\cos \gamma_{1}}-\frac{\cos\ag_{2}+ i \cos\bg_{2}}{1+\cos\gamma_{2}}\right],\\
		&& \Omega=\frac{1}{2\sqrt{2}}\sqrt{(1+\cos \gamma_{1})(1+\cos \gamma_{2})}=N_1N_2={\cal N}_{1,2}.~~~~~~~
	\end{eqnarray}
\end{widetext}

For $N=3$ case we introduce few symbols to write the angular dependence of the spectral shift\\

\begin{widetext}
	\begin{eqnarray}
		&& \Omega_1=\frac{1}{2} \sqrt{1+\cos \gamma_{1}},\\
		\displaystyle
		&& \Omega_2=\frac{1}{2} \sqrt{1+\cos \gamma_{2}},\\
		\displaystyle
		&& \Omega_3=\frac{1}{2} \sqrt{1+\cos \gamma_{3}},\\
		\displaystyle
		&& \alpha_{12}=\cos \alpha_1 \cos \alpha_2,\\
		\displaystyle
		&& \beta_{12}=\cos \beta_1 \cos \beta_2,
		\\
		\displaystyle
		&& \tilde \alpha_1=\cos \alpha_1 - i \cos \beta_1,
	\end{eqnarray}
	\begin{eqnarray}
		\displaystyle
		&& \tilde \alpha_2=\cos \alpha_2 - i \cos \beta_2,\\
		\displaystyle
		&& \tilde \alpha_3=\cos \alpha_3 - i \cos \beta_3,\\
		\displaystyle
		&& \tilde \alpha_1^*=\cos \alpha_1 + i \cos \beta_1,\\
		\displaystyle
		&& \tilde \alpha_2^*=\cos \alpha_2 + i \cos \beta_2,\\
		\displaystyle
		&& \tilde \alpha_3^*=\cos \alpha_3 + i \cos \beta_3
	\end{eqnarray}
\end{widetext}
Therefore the angular dependence for the ground state in this case can be written as:
\begin{equation}
	\Gamma_{1;{\cal DC}}=\frac{1}{6}(\mathcal{G}_1+\mathcal{G}_2+\mathcal{G}_3+\mathcal{G}_4)
\end{equation}
where we define:
\begin{eqnarray}
	&&\mathcal{G}_1=2( \Omega_1 \Omega_2 +  \Omega_1 \Omega_3 +  \Omega_2 \Omega_3 ) \nonumber\\
	&&~~~~~\left[-2 i (\alpha_{12}- \beta_{12}) \left(\frac{\tilde{\alpha_1} \tilde{\alpha_2}}{6\Omega_1 \Omega_2} + \frac{\tilde{\alpha_1} \tilde{\alpha_3}}{6\Omega_1 \Omega_3}  + \frac{\tilde{\alpha_2} \tilde{\alpha_3}}{6\Omega_2 \Omega_3} \right)
	\right.\nonumber\\ &&\left.~~~~~~ +2 i (\alpha_{12} +i \beta_{12}) \left(\frac{\tilde{\alpha_2} \Omega_1}{2 \Omega_2} + \frac{\tilde{\alpha_3} \Omega_2}{2 \Omega_3}  + \frac{\tilde{\alpha_1} \Omega_3}{2\Omega_1} \right)  \right.\nonumber\\ &&\left.~~~~~~
	+2 i (\alpha_{12} +i \beta_{12}) \left(\frac{\tilde{\alpha_1} \Omega_2}{2 \Omega_1} + \frac{\tilde{\alpha_3} \Omega_1}{2 \Omega_3}  + \frac{\tilde{\alpha_2} \Omega_3}{2\Omega_2} \right) \right],~~~~
\end{eqnarray}
\begin{eqnarray}
	&&\mathcal{G}_2=\left(\frac{\tilde{\alpha_1}^* \tilde{\alpha_2}^*}{6 \Omega_1 \Omega_2} + \frac{\tilde{\alpha_1}^* \tilde{\alpha_3}^*}{6 \Omega_1 \Omega_3 }+ \frac{\tilde{\alpha_2}^* \tilde{\alpha_3}^*}{6 \Omega_2 \Omega_3}\right) \nonumber\\
	&&~~~~~ \left[2 i(\alpha_{12}-i \beta_{12}) \left(\frac{\tilde{\alpha_2} \Omega_1}{2 \Omega_2} +  \frac{\tilde{\alpha_3} \Omega_2}{2 \Omega_3}+ \frac{\tilde{\alpha_1} \Omega_3}{2 \Omega_1}\right)\right.\nonumber\\ &&\left.~~~~~~+2 i(\alpha_{12}-i \beta_{12})\left(\frac{\tilde{\alpha_1} \Omega_2}{2 \Omega_1} +\frac{\tilde{\alpha_3} \Omega_1}{2 \Omega_3}+ \frac{\tilde{\alpha_2} \Omega_3}{2 \Omega_2} \right)
	\right. \nonumber\\ &&\left.~~~~~~ -2 i(\alpha_{12}- \beta_{12})(2 \Omega_1 \Omega_2 + 2 \Omega_1 \Omega_3 +2 \Omega_2 \Omega_3) \right],~~~~
\end{eqnarray}
\begin{eqnarray}
	&&\mathcal{G}_3=\left(\frac{\tilde{\alpha_1}^* \Omega_2}{2 \Omega_1 } - \frac{\tilde{\alpha_3}^* \Omega_1}{2 \Omega_3 }- \frac{\tilde{\alpha_2}^* \Omega_3}{2 \Omega_2 }\right) \nonumber\\
	&&~~~~~  \left[-2 i (\alpha_{12}+ i \beta_{12}) \left(\frac{\tilde{\alpha_1} \tilde{\alpha_2}}{6 \Omega_1 \Omega_2} + \frac{\tilde{\alpha_1} \tilde{\alpha_3}}{6 \Omega_1 \Omega_3}+ \frac{\tilde{\alpha_2} \tilde{\alpha_3}}{6 \Omega_2 \Omega_3}\right)\right.\nonumber\\ &&\left.~~~~~~  + 2 i (\alpha_{12}+\beta_{12})\left(\frac{\tilde{\alpha_2} \Omega_1}{2 \Omega_2} 
	+\frac{\tilde{\alpha_3} \Omega_2}{2 \Omega_3}+ \frac{\tilde{\alpha_1} \Omega_3}{2 \Omega_1} \right)
	\right.\nonumber\\ &&\left.~~~~~~ 
	-2 i(\alpha_{12}-i \beta_{12}) \left(2 \Omega_1 \Omega_2 + 2 \Omega_1 \Omega_3 +2 \Omega_2 \Omega_3\right) \right],~~~~
\end{eqnarray}
\begin{eqnarray}
	&&\mathcal{G}_4=-\left(\frac{\tilde{\alpha_2}^* \Omega_1}{2 \Omega_2 } + \frac{\tilde{\alpha_3}^* \Omega_2}{2 \Omega_3 }+ \frac{\tilde{\alpha_1}^* \Omega_3}{2 \Omega_1 }\right) \nonumber\\
	&&~~~~~   \left[-2 i (\alpha_{12}+ i \beta_{12}) \left(\frac{\tilde{\alpha_1} \tilde{\alpha_2}}{6 \Omega_1 \Omega_2} + \frac{\tilde{\alpha_1} \tilde{\alpha_3}}{6 \Omega_1 \Omega_3}+ \frac{\tilde{\alpha_2} \tilde{\alpha_3}}{6 \Omega_2 \Omega_3}\right)\right.\nonumber\\ &&\left.~~~~~~  - 2 i (\alpha_{12}+\beta_{12})\left(\frac{\tilde{\alpha_1} \Omega_2}{2 \Omega_1} 
	+\frac{\tilde{\alpha_3} \Omega_1}{2 \Omega_3}+ \frac{\tilde{\alpha_2} \Omega_3}{2 \Omega_2} \right)
	\right.\nonumber\\ &&\left.~~~~~~  -2 i(\alpha_{12}-i \beta_{12}) \left(2 \Omega_1 \Omega_2 + 2 \Omega_1 \Omega_3 +2 \Omega_2 \Omega_3\right) \right].~~~~
\end{eqnarray}
Therefore the angular dependence for the excited state in this case can be written as:
\begin{equation}
	\Gamma_{1;{\cal DC}}=\frac{1}{6}(\mathcal{E}_1+\mathcal{E}_2+\mathcal{E}_3+\mathcal{E}_4)
\end{equation}
where we define:
\begin{eqnarray}
	&&\mathcal{E}_1=2( \Omega_1 \Omega_2 +  \Omega_1 \Omega_3 +  \Omega_2 \Omega_3 ) \nonumber\\
	&&~~~~~  \left[-2 i (\alpha_{12}- \beta_{12}) \left(\frac{\tilde{\alpha_1}^* \tilde{\alpha_2}^*}{6\Omega_1 \Omega_2} + \frac{\tilde{\alpha_1}^* \tilde{\alpha_3}^*}{6\Omega_1 \Omega_3}  + \frac{\tilde{\alpha_2}^* \tilde{\alpha_3}^*}{6\Omega_2 \Omega_3} \right)
	\right.\nonumber\\ &&\left.~~~~~~  -2 i (\alpha_{12} -i \beta_{12}) \left(\frac{\tilde{\alpha_2}^* \Omega_1}{2 \Omega_2} + \frac{\tilde{\alpha_3}^* \Omega_2}{2 \Omega_3}  + \frac{\tilde{\alpha_1}^* \Omega_3}{2\Omega_1} \right)  \right.\nonumber\\ &&\left.~~~~~~ 
	-2 i (\alpha_{12} -i \beta_{12}) \left(\frac{\tilde{\alpha_1}^* \Omega_2}{2 \Omega_1} + \frac{\tilde{\alpha_3}^* \Omega_1}{2 \Omega_3}  + \frac{\tilde{\alpha_2}^* \Omega_3}{2\Omega_2} \right) \right],~~~~~~
\end{eqnarray}
\begin{eqnarray}
	&&\mathcal{E}_2=\left(\frac{\tilde{\alpha_1} \tilde{\alpha_2}}{6 \Omega_1 \Omega_2} + \frac{\tilde{\alpha_1} \tilde{\alpha_3}}{6 \Omega_1 \Omega_3 }+ \frac{\tilde{\alpha_2} \tilde{\alpha_3}}{6 \Omega_2 \Omega_3}\right)  \nonumber\\
	&&~~~~~\left[-2 i(\alpha_{12}+i \beta_{12}) \left(\frac{\tilde{\alpha_2}^* \Omega_1}{2 \Omega_2} +  \frac{\tilde{\alpha_3}^* \Omega_2}{2 \Omega_3}+ \frac{\tilde{\alpha_1}^* \Omega_3}{2 \Omega_1}\right)  \right.\nonumber\\ &&\left.~~~~~~ -2 i(\alpha_{12}+i \beta_{12})\left(\frac{\tilde{\alpha_1}^* \Omega_2}{2 \Omega_1} +\frac{\tilde{\alpha_3}^* \Omega_1}{2 \Omega_3}+ \frac{\tilde{\alpha_2}^* \Omega_3}{2 \Omega_2} \right)
	\right. \nonumber\\ &&\left.~~~~~~ -2 i(\alpha_{12}- \beta_{12})(2 \Omega_1 \Omega_2 + 2 \Omega_1 \Omega_3 +2 \Omega_2 \Omega_3) \right],~~~~~~
\end{eqnarray}
\begin{eqnarray}
	&&\mathcal{E}_3=\left(\frac{\tilde{\alpha_1} \Omega_2}{2 \Omega_1 } - \frac{\tilde{\alpha_3} \Omega_1}{2 \Omega_3 }- \frac{\tilde{\alpha_2} \Omega_3}{2 \Omega_2 }\right)\nonumber\\
	&&~~~~~  \left[-2 i (\alpha_{12}- i \beta_{12}) \left(\frac{\tilde{\alpha_1}^* \tilde{\alpha_2}^*}{6 \Omega_1 \Omega_2} + \frac{\tilde{\alpha_1}^* \tilde{\alpha_3}^*}{6 \Omega_1 \Omega_3}+ \frac{\tilde{\alpha_2}^* \tilde{\alpha_3}^*}{6 \Omega_2 \Omega_3}\right)\right.\nonumber\\ &&\left.~~~~~~  - 2 i (\alpha_{12}+\beta_{12})\left(\frac{\tilde{\alpha_2}^* \Omega_1}{2 \Omega_2} 
	+\frac{\tilde{\alpha_3}^* \Omega_2}{2 \Omega_3}+ \frac{\tilde{\alpha_1}^* \Omega_3}{2 \Omega_1} \right)
	\right.\nonumber\\ &&\left.~~~~~~  
	-2 i(\alpha_{12}+i \beta_{12}) \left(2 \Omega_1 \Omega_2 + 2 \Omega_1 \Omega_3 +2 \Omega_2 \Omega_3\right) \right],~~~~~~
\end{eqnarray}
\begin{eqnarray}
	&&\mathcal{E}_4=\left(\frac{\tilde{\alpha_2} \Omega_1}{2 \Omega_2 } + \frac{\tilde{\alpha_3} \Omega_2}{2 \Omega_3 }+ \frac{\tilde{\alpha_1} \Omega_3}{2 \Omega_1 }\right)\nonumber\\
	&&~~~~~  \left[-2 i (\alpha_{12}- i \beta_{12}) \left(\frac{\tilde{\alpha_1}^* \tilde{\alpha_2}^*}{6 \Omega_1 \Omega_2} + \frac{\tilde{\alpha_1}^* \tilde{\alpha_3}^*}{6 \Omega_1 \Omega_3}+ \frac{\tilde{\alpha_2}^* \tilde{\alpha_3}^*}{6 \Omega_2 \Omega_3}\right)\right.\nonumber\\ &&\left.~~~~~~ - 2 i (\alpha_{12}+\beta_{12})\left(\frac{\tilde{\alpha_1}^* \Omega_2}{2 \Omega_1} 
	+\frac{\tilde{\alpha_3}^* \Omega_1}{2 \Omega_3}+ \frac{\tilde{\alpha_2}^* \Omega_3}{2 \Omega_2} \right)
	\right.\nonumber\\ &&\left.~~~~~~ -2 i(\alpha_{12}+i \beta_{12}) \left(2 \Omega_1 \Omega_2 + 2 \Omega_1 \Omega_3 +2 \Omega_2 \Omega_3\right) \right],~~~~~~
\end{eqnarray}
Therefore the angular dependence for the Symmetric state in this case can be written as:
\begin{equation}
	\Gamma_{2;{\cal DC}}=\frac{1}{6}(\mathcal{S}_1+\mathcal{S}_2+\mathcal{S}_3+\mathcal{S}_4)
\end{equation}
where we define:
\begin{eqnarray}
	&& \mathcal{S}_1=\left(-\frac{ \tilde{\alpha_1}^* \tilde{\alpha_2}}{6 \Omega_1 \Omega_2} - \frac{ \tilde{\alpha_1}^* \tilde{\alpha_3}}{6 \Omega_1 \Omega_3}- \frac{ \tilde{\alpha_2}^* \tilde{\alpha_3}}{6 \Omega_2 \Omega_3}  + 2 \Omega_1 \Omega_2 + 2 \Omega_1 \Omega_3 + 2 \Omega_2 \Omega_3 \right) \nonumber\\ 
	&&~~~~~~~~~~~~~ \left[- i(\alpha_{12}+i \beta_{12}) \left(-\frac{\tilde{\alpha_2} \Omega_1}{2 \Omega_2} - \frac{\tilde{\alpha_1} \Omega_2}{2 \Omega_1}- \frac{\tilde{\alpha_3} \Omega_1}{2 \Omega_3}\right.\right.\nonumber\\ &&\left.\left.~~~~~~~~~~~~~~~~~~~~~~ -\frac{\tilde{\alpha_3} \Omega_2}{2 \Omega_3}-\frac{\tilde{\alpha_1} \Omega_3}{2 \Omega_1}-\frac{\tilde{\alpha_2} \Omega_3}{2 \Omega_2} \right) \right.\nonumber\\ &&\left.~~~~~~ 
	- i (\alpha_{12} -i \beta_{12}) \left(\frac{\tilde{\alpha_2} \Omega_1}{2 \Omega_2} + \frac{\tilde{\alpha_1} \Omega_2}{2 \Omega_1}  + \frac{\tilde{\alpha_1} \Omega_3}{2\Omega_1}\right.\right.\nonumber\\ &&\left.\left.~~~~~~~~~~~~~~~~~~~~~~+ \frac{\tilde{\alpha_3} \Omega_1}{2 \Omega_3} +\frac{\tilde{\alpha_3} \Omega_2}{2 \Omega_3} +\frac{\tilde{\alpha_2} \Omega_3}{2 \Omega_2} \right)  \right.\nonumber\\ &&\left.~~~~~~ 
	- i (\alpha_{12} + \beta_{12}) \left(-\frac{\tilde{\alpha_1}^* \tilde{\alpha_2} }{6 \Omega_1 \Omega_2}  -  \frac{\tilde{\alpha_2}^* \tilde{\alpha_3}}{6 \Omega_2 \Omega_3}- \frac{\tilde{\alpha_1}^* \tilde{\alpha_3}}{6 \Omega_1 \Omega_3} \right.\right.\nonumber\\ &&\left.\left.~~~~~~~~~~~~~~~~~~~~~~+ 2 \Omega_1 \Omega_2 +2 \Omega_1 \Omega_3 +2 \Omega_2 \Omega_3 \right) \right],~~~~~~
	\\
	&&\mathcal{S}_2=	\left(-\frac{\tilde{\alpha_2}^* \Omega_1}{2 \Omega_2} - \frac{\tilde{\alpha_1}^* \Omega_2}{2 \Omega_1}- \frac{\tilde{\alpha_3}^* \Omega_1}{2 \Omega_3}\right.\nonumber\\ &&\left.~~~~~~~~~~~~~~~~~~~~~~-\frac{\tilde{\alpha_3}^* \Omega_2}{2 \Omega_3}-\frac{\tilde{\alpha_1}^* \Omega_3}{2 \Omega_1}-\frac{\tilde{\alpha_2}^* \Omega_3}{2 \Omega_2}\right) \nonumber\\ 
	&&~~~~~~~~~~~~~ \left[
	- i (\alpha_{12} - \beta_{12}) \left(\frac{\tilde{\alpha_2}^* \Omega_1}{2 \Omega_2} + \frac{\tilde{\alpha_1}^* \Omega_2}{2 \Omega_1}  + \frac{\tilde{\alpha_3}^* \Omega_1}{2\Omega_3}\right.\right.\nonumber\\ &&\left.\left.~~~~~~~~~~~~~~~~~~~~~~+ \frac{\tilde{\alpha_3}^* \Omega_2}{2 \Omega_3} +\frac{\tilde{\alpha_2}^* \Omega_3}{2 \Omega_2} +\frac{\tilde{\alpha_1}^* \Omega_3}{2 \Omega_1} \right) \right.\nonumber\\ &&\left.~~~~~~ 
	- i (\alpha_{12} -i \beta_{12}) \left(-\frac{\tilde{\alpha_1}^* \tilde{\alpha_2} }{6 \Omega_1 \Omega_2}  -  \frac{\tilde{\alpha_1}^* \tilde{\alpha_3}}{6 \Omega_1 \Omega_3}- \frac{\tilde{\alpha_2}^* \tilde{\alpha_3}}{6 \Omega_2 \Omega_3}\right.\right.\nonumber\\ &&\left.\left.~~~~~~~~~~~~~~~~~~~~~~  + 2 \Omega_1 \Omega_2 +2 \Omega_1 \Omega_3 +2 \Omega_2 \Omega_3 \right)  \right.\nonumber\\ &&\left.~~~~~~ 
	- i (\alpha_{12} -i \beta_{12}) \left(-\frac{\tilde{\alpha_1} \tilde{\alpha_2}^* }{6 \Omega_1 \Omega_2}  -  \frac{\tilde{\alpha_1} \tilde{\alpha_3}^*}{6 \Omega_1 \Omega_3}- \frac{\tilde{\alpha_2} \tilde{\alpha_3}^*}{6 \Omega_2 \Omega_3} \right.\right.\nonumber\\ &&\left.\left.~~~~~~~~~~~~~~~~~~~~~~ + 2 \Omega_1 \Omega_2 +2 \Omega_1 \Omega_3 +2 \Omega_2 \Omega_3 \right)
	\right]
	\\
	&&\mathcal{S}_3=	\left(\frac{\tilde{\alpha_2} \Omega_1}{2 \Omega_2} + \frac{\tilde{\alpha_1} \Omega_2}{2 \Omega_1}+ \frac{\tilde{\alpha_3} \Omega_1}{2 \Omega_3}+\frac{\tilde{\alpha_3} \Omega_2}{2 \Omega_3}+\frac{\tilde{\alpha_1} \Omega_3}{2 \Omega_1}+\frac{\tilde{\alpha_2} \Omega_3}{2 \Omega_2}\right) \nonumber\\ 
	&&~~~~~~~~~~~~~  \left[
	- i (\alpha_{12} - \beta_{12}) \left(-\frac{\tilde{\alpha_2} \Omega_1}{2 \Omega_2} -\frac{\tilde{\alpha_1} \Omega_2}{2 \Omega_1}  - \frac{\tilde{\alpha_3} \Omega_1}{2\Omega_3}\right.\right.\nonumber\\ &&\left.\left.~~~~~~~~~~~~~~~~~~~~~~- \frac{\tilde{\alpha_3} \Omega_2}{2 \Omega_3} -\frac{\tilde{\alpha_2} \Omega_3}{2 \Omega_2} -\frac{\tilde{\alpha_1} \Omega_3}{2 \Omega_1} \right)   \right.\nonumber\\ &&\left.~~~~~~ 
	- i (\alpha_{12} +i \beta_{12}) \left(-\frac{\tilde{\alpha_1}^* \tilde{\alpha_2} }{6 \Omega_1 \Omega_2}  -  \frac{\tilde{\alpha_1}^* \tilde{\alpha_3}}{6 \Omega_1 \Omega_3}- \frac{\tilde{\alpha_2}^* \tilde{\alpha_3}}{6 \Omega_2 \Omega_3}\right.\right.\nonumber\\ &&\left.\left.~~~~~~~~~~~~~~~~~~~~~~  + 2 \Omega_1 \Omega_2 +2 \Omega_1 \Omega_3 +2 \Omega_2 \Omega_3 \right)   \right.\nonumber\\ &&\left.~~~~~~ 
	- i (\alpha_{12} +i \beta_{12}) \left(-\frac{\tilde{\alpha_1} \tilde{\alpha_2}^* }{6 \Omega_1 \Omega_2}  -  \frac{\tilde{\alpha_1} \tilde{\alpha_3}^*}{6 \Omega_1 \Omega_3}- \frac{\tilde{\alpha_2} \tilde{\alpha_3}^*}{6 \Omega_2 \Omega_3}  \right.\right.\nonumber\\ &&\left.\left.~~~~~~~~~~~~~~~~~~~~~~+ 2 \Omega_1 \Omega_2 +2 \Omega_1 \Omega_3 +2 \Omega_2 \Omega_3 \right)
	\right]
	\eea\bea
	&&\mathcal{S}_4=\left(-\frac{ \tilde{\alpha_1} \tilde{\alpha_2}^*}{6 \Omega_1 \Omega_2} - \frac{ \tilde{\alpha_1} \tilde{\alpha_3}^*}{6 \Omega_1 \Omega_3}- \frac{ \tilde{\alpha_2} \tilde{\alpha_3}^*}{6 \Omega_2 \Omega_3}  \right.\nonumber\\ &&\left.~~~~~~~~~~~~~~~~~~~~~~+ 2 \Omega_1 \Omega_2 + 2 \Omega_1 \Omega_3 + 2 \Omega_2 \Omega_3 \right) \nonumber\\ 
	&&~~~~~~~~~~~~~  \left	[- i(\alpha_{12}+i \beta_{12}) \left(-\frac{\tilde{\alpha_2} \Omega_1}{2 \Omega_2} - \frac{\tilde{\alpha_1} \Omega_2}{2 \Omega_1}- \frac{\tilde{\alpha_3} \Omega_1}{2 \Omega_3}\right.\right.\nonumber\\ &&\left.\left.~~~~~~~~~~~~~~~~~~~~~~-\frac{\tilde{\alpha_3} \Omega_2}{2 \Omega_3}-\frac{\tilde{\alpha_1} \Omega_3}{2 \Omega_1}-\frac{\tilde{\alpha_2} \Omega_3}{2 \Omega_2} \right)  \right.\nonumber\\ &&\left.~~~~~~ 
	- i (\alpha_{12} -i \beta_{12}) \left(\frac{\tilde{\alpha_2}^* \Omega_1}{2 \Omega_2} + \frac{\tilde{\alpha_1}^* \Omega_2}{2 \Omega_3}  + \frac{\tilde{\alpha_3}^* \Omega_1}{2\Omega_3}\right.\right.\nonumber\\ &&\left.\left.~~~~~~~~~~~~~~~~~~~~~~+ \frac{\tilde{\alpha_3}^* \Omega_2}{2 \Omega_3} +\frac{\tilde{\alpha_2}^* \Omega_3}{2 \Omega_2} +\frac{\tilde{\alpha_1}^* \Omega_3}{2 \Omega_1} \right)  \right.\nonumber\\ &&\left.~~~~~~ 
	- i (\alpha_{12} + \beta_{12}) \left(-\frac{\tilde{\alpha_1} \tilde{\alpha_2}^* }{6 \Omega_1 \Omega_2}  -  \frac{\tilde{\alpha_2} \tilde{\alpha_3}^*}{6 \Omega_2 \Omega_3}- \frac{\tilde{\alpha_1} \tilde{\alpha_3}^*}{6 \Omega_1 \Omega_3} \right.\right.\nonumber\\ &&\left.\left.~~~~~~~~~~~~~~~~~~~~~~+ 2 \Omega_1 \Omega_2 +2 \Omega_1 \Omega_3 +2 \Omega_2 \Omega_3 \right) \right]
\end{eqnarray}
Therefore,the angular dependence for the Antisymmetric state in this case can be written as:
\begin{equation}
	\Gamma_{3;{\cal DC}}=\frac{1}{6}(\mathcal{A}_1+\mathcal{A}_2+\mathcal{A}_3+\mathcal{A}_4)
\end{equation}
where we define:
\begin{eqnarray}
	&&\mathcal{A}_1=\left(\frac{ \tilde{\alpha_1}^* \tilde{\alpha_2}}{6 \Omega_1 \Omega_2} + \frac{ \tilde{\alpha_1}^* \tilde{\alpha_3}}{6 \Omega_1 \Omega_3}+ \frac{ \tilde{\alpha_2}^* \tilde{\alpha_3}}{6 \Omega_2 \Omega_3} + 2 \Omega_1 \Omega_2 + 2 \Omega_1 \Omega_3 + 2 \Omega_2 \Omega_3 \right) \nonumber\\ 
	&&~~~~~~~~~~~~~ \left	[- i(\alpha_{12}+i \beta_{12}) \left(-\frac{\tilde{\alpha_2} \Omega_1}{2 \Omega_2} + \frac{\tilde{\alpha_1} \Omega_2}{2 \Omega_1}- \frac{\tilde{\alpha_3} \Omega_1}{2 \Omega_3}\right.\right.\nonumber\\ &&\left.\left.~~~~~~~~~~~~~~~~~~~~~~-\frac{\tilde{\alpha_3} \Omega_2}{2 \Omega_3}+\frac{\tilde{\alpha_1} \Omega_3}{2 \Omega_1}+\frac{\tilde{\alpha_2} \Omega_3}{2 \Omega_2} \right)  \right.\nonumber\\ &&\left.~~~~~~ 
	- i (\alpha_{12} -i \beta_{12}) \left(-\frac{\tilde{\alpha_2} \Omega_1}{2 \Omega_2} + \frac{\tilde{\alpha_1} \Omega_2}{2 \Omega_1}  + \frac{\tilde{\alpha_1} \Omega_3}{2\Omega_1}\right.\right.\nonumber\\ &&\left.\left.~~~~~~~~~~~~~~~~~~~~~~- \frac{\tilde{\alpha_3} \Omega_1}{2 \Omega_3} -\frac{\tilde{\alpha_3} \Omega_2}{2 \Omega_3} +\frac{\tilde{\alpha_2} \Omega_3}{2 \Omega_2} \right)  \right.\nonumber\\ &&\left.~~~~~~ 
	- i (\alpha_{12} + \beta_{12}) \left(-\frac{\tilde{\alpha_1}^* \tilde{\alpha_2} }{6 \Omega_1 \Omega_2}  -  \frac{\tilde{\alpha_2}^* \tilde{\alpha_3}}{6 \Omega_2 \Omega_3}- \frac{\tilde{\alpha_1}^* \tilde{\alpha_3}}{6 \Omega_1 \Omega_3}\right.\right.\nonumber\\ &&\left.\left.~~~~~~~~~~~~~~~~~~~~~~ - 2 \Omega_1 \Omega_2 -2 \Omega_1 \Omega_3 -2 \Omega_2 \Omega_3 \right) \right],~~~~~~
\end{eqnarray}
\bea
&&\mathcal{A}_2=	\left(-\frac{\tilde{\alpha_2}^* \Omega_1}{2 \Omega_2} + \frac{\tilde{\alpha_1}^* \Omega_2}{2 \Omega_1}- \frac{\tilde{\alpha_3}^* \Omega_1}{2 \Omega_3} -\frac{\tilde{\alpha_3}^* \Omega_2}{2 \Omega_3}+\frac{\tilde{\alpha_1}^* \Omega_3}{2 \Omega_1}+\frac{\tilde{\alpha_2}^* \Omega_3}{2 \Omega_2}\right) \nonumber\\ 
&&~~~~~~~~~~~~~ \left[
- i (\alpha_{12} - \beta_{12}) \left(-\frac{\tilde{\alpha_2}^* \Omega_1}{2 \Omega_2} + \frac{\tilde{\alpha_1}^* \Omega_2}{2 \Omega_1}  - \frac{\tilde{\alpha_3}^* \Omega_1}{2\Omega_3}\right.\right.\nonumber\\ &&\left.\left.~~~~~~~~~~~~~~~~~~~~~~ - \frac{\tilde{\alpha_3}^* \Omega_2}{2 \Omega_3} +\frac{\tilde{\alpha_2}^* \Omega_3}{2 \Omega_2} +\frac{\tilde{\alpha_1}^* \Omega_3}{2 \Omega_1} \right)  \right.\nonumber\\ &&\left.~~~~~~ 
- i (\alpha_{12} -i \beta_{12}) \left(-\frac{\tilde{\alpha_1}^* \tilde{\alpha_2} }{6 \Omega_1 \Omega_2}  -  \frac{\tilde{\alpha_1}^* \tilde{\alpha_3}}{6 \Omega_1 \Omega_3}- \frac{\tilde{\alpha_2}^* \tilde{\alpha_3}}{6 \Omega_2 \Omega_3} \right.\right.\nonumber\\ &&\left.\left.~~~~~~~~~~~~~~~~~~~~~~  - 2 \Omega_1 \Omega_2 -2 \Omega_1 \Omega_3 -2 \Omega_2 \Omega_3 \right)  \right.\nonumber\\ &&\left.~~~~~~ 
- i (\alpha_{12} -i \beta_{12}) \left(\frac{\tilde{\alpha_1} \tilde{\alpha_2}^* }{6 \Omega_1 \Omega_2}  +  \frac{\tilde{\alpha_1} \tilde{\alpha_3}^*}{6 \Omega_1 \Omega_3} + \frac{\tilde{\alpha_2} \tilde{\alpha_3}^*}{6 \Omega_2 \Omega_3} \right.\right.\nonumber\\ &&\left.\left.~~~~~~~~~~~~~~~~~~~~~~  + 2 \Omega_1 \Omega_2 +2 \Omega_1 \Omega_3 +2 \Omega_2 \Omega_3 \right)
\right],~~~~~~
\eea
\bea
&&~~~~~~~~~~\mathcal{A}_3=	\left(-\frac{\tilde{\alpha_2} \Omega_1}{2 \Omega_2} + \frac{\tilde{\alpha_1} \Omega_2}{2 \Omega_1}- \frac{\tilde{\alpha_3} \Omega_1}{2 \Omega_3}-\frac{\tilde{\alpha_3} \Omega_2}{2 \Omega_3}+\frac{\tilde{\alpha_1} \Omega_3}{2 \Omega_1}+\frac{\tilde{\alpha_2} \Omega_3}{2 \Omega_2}\right) \nonumber\\ 
&&~~~~~~~~~~~~~ \left[
- i (\alpha_{12} - \beta_{12}) \left(-\frac{\tilde{\alpha_2} \Omega_1}{2 \Omega_2} +\frac{\tilde{\alpha_1} \Omega_2}{2 \Omega_1}  - \frac{\tilde{\alpha_3} \Omega_1}{2\Omega_3}\right.\right.\nonumber\\ &&\left.\left.~~~~~~~~~~~~~~~~~~~~~~  - \frac{\tilde{\alpha_3} \Omega_2}{2 \Omega_3} +\frac{\tilde{\alpha_2} \Omega_3}{2 \Omega_2} +\frac{\tilde{\alpha_1} \Omega_3}{2 \Omega_1} \right)  \right.\nonumber\\ &&\left.~~~~~~ 
- i (\alpha_{12} +i \beta_{12}) \left(-\frac{\tilde{\alpha_1}^* \tilde{\alpha_2} }{6 \Omega_1 \Omega_2}  -  \frac{\tilde{\alpha_1}^* \tilde{\alpha_3}}{6 \Omega_1 \Omega_3}- \frac{\tilde{\alpha_2}^* \tilde{\alpha_3}}{6 \Omega_2 \Omega_3}\right.\right.\nonumber\\ &&\left.\left.~~~~~~~~~~~~~~~~~~~~~~    - 2 \Omega_1 \Omega_2 -2 \Omega_1 \Omega_3 - 2 \Omega_2 \Omega_3 \right)  \right.\nonumber\\ &&\left.~~~~~~ 
- i (\alpha_{12} +i \beta_{12}) \left(\frac{\tilde{\alpha_1} \tilde{\alpha_2}^* }{6 \Omega_1 \Omega_2}  +  \frac{\tilde{\alpha_1} \tilde{\alpha_3}^*}{6 \Omega_1 \Omega_3}+ \frac{\tilde{\alpha_2} \tilde{\alpha_3}^*}{6 \Omega_2 \Omega_3} \right.\right.\nonumber\\ &&\left.\left.~~~~~~~~~~~~~~~~~~~~~~   + 2 \Omega_1 \Omega_2 +2 \Omega_1 \Omega_3 +2 \Omega_2 \Omega_3 \right)
\right],~~~~~~
\eea
\bea
&&\mathcal{A}_4=\left(-\frac{ \tilde{\alpha_1} \tilde{\alpha_2}^*}{6 \Omega_1 \Omega_2} - \frac{ \tilde{\alpha_1} \tilde{\alpha_3}^*}{6 \Omega_1 \Omega_3}- \frac{ \tilde{\alpha_2} \tilde{\alpha_3}^*}{6 \Omega_2 \Omega_3}   - 2 \Omega_1 \Omega_2 - 2 \Omega_1 \Omega_3 - 2 \Omega_2 \Omega_3 \right) \nonumber\\ &&~~~~~~  \left	[- i(\alpha_{12}+i \beta_{12}) \left(-\frac{\tilde{\alpha_2} \Omega_1}{2 \Omega_2} + \frac{\tilde{\alpha_1} \Omega_2}{2 \Omega_1}- \frac{\tilde{\alpha_3} \Omega_1}{2 \Omega_3} \right.\right.\nonumber\\ &&\left.\left.~~~~~~~~~~~~~~~~~~~~~~-\frac{\tilde{\alpha_3} \Omega_2}{2 \Omega_3}+\frac{\tilde{\alpha_1} \Omega_3}{2 \Omega_1}+\frac{\tilde{\alpha_2} \Omega_3}{2 \Omega_2} \right)  \right.\nonumber\\ &&\left.~~~~~~ 
- i (\alpha_{12} -i \beta_{12}) \left(\frac{-\tilde{\alpha_2}^* \Omega_1}{2 \Omega_2} + \frac{\tilde{\alpha_1}^* \Omega_2}{2 \Omega_3}  - \frac{\tilde{\alpha_3}^* \Omega_1}{2\Omega_3} \right.\right.\nonumber\\ &&\left.\left.~~~~~~~~~~~~~~~~~~~~~~- \frac{\tilde{\alpha_3}^* \Omega_2}{2 \Omega_3} +\frac{\tilde{\alpha_2}^* \Omega_3}{2 \Omega_2} +\frac{\tilde{\alpha_1}^* \Omega_3}{2 \Omega_1} \right)  \right.\nonumber\\ &&\left.~~~~~~ 
- i (\alpha_{12} + \beta_{12}) \left(\frac{\tilde{\alpha_1} \tilde{\alpha_2}^* }{6 \Omega_1 \Omega_2}  +  \frac{\tilde{\alpha_2} \tilde{\alpha_3}^*}{6 \Omega_2 \Omega_3}+ \frac{\tilde{\alpha_1} \tilde{\alpha_3}^*}{6 \Omega_1 \Omega_3}  \right.\right.\nonumber\\ &&\left.\left.~~~~~~~~~~~~~~~~~~~~~~+ 2 \Omega_1 \Omega_2 +2 \Omega_1 \Omega_3 +2 \Omega_2 \Omega_3 \right) \right].~~~~~~
\end{eqnarray}
\newpage
{\section{E. Spectroscopic shifts for $N$ spins in static patch of de Sitter space}
\label{Appendix:E}}
To compute the spectroscopic shifts from the entangled ground, excited, symmetric and antisymmetric states we need to compute the following expressions for $N$ spin system:
\begin{widetext}
\bea &&\textcolor{blue}{\bf Ground~~state:}~~~~~\delta E^{N}_{G}=\langle G|H_{LS}|G\rangle =-\frac{i}{2}\sum^{N}_{\delta,\eta=1}\sum^{3}_{i,j=1}H^{(\delta\eta)}_{ij} \langle G|(n^{\delta}_{i}.\sigma^{\delta}_{i})(n^{\eta}_{j}.\sigma^{\eta}_{j})| G\rangle=-\frac{2P{\cal F}(L,k,\omega_0)\Gamma^{N}_{1;{\cal DC}}}{{\cal N}^2_{\rm norm}},\\
&&\textcolor{blue}{\bf Excited~~state:}~~~~~\delta E^{N}_{E}=\langle E|H_{LS}|E\rangle =-\frac{i}{2}\sum^{N}_{\delta,\eta=1}\sum^{3}_{i,j=1}H^{(\delta\eta)}_{ij} \langle E|(n^{\delta}_{i}.\sigma^{\delta}_{i})(n^{\eta}_{j}.\sigma^{\eta}_{j})| E\rangle=-\frac{2P{\cal F}(L,k,\omega_0)\Gamma^{N}_{1;{\cal DC}}}{{\cal N}^2_{\rm norm}},\\
&&\textcolor{blue}{\bf Symmetric~~state:}~~~~~\delta E^{N}_{S}=\langle S|H_{LS}|S\rangle =-\frac{i}{2}\sum^{N}_{\delta,\eta=1}\sum^{3}_{i,j=1}H^{(\delta\eta)}_{ij} \langle S|(n^{\delta}_{i}.\sigma^{\delta}_{i})(n^{\eta}_{j}.\sigma^{\eta}_{j})| S\rangle=-\frac{P{\cal F}(L,k,\omega_0)\Gamma^{N}_{2;{\cal DC}}}{{\cal N}^2_{\rm norm}},,\\
&&\textcolor{blue}{\bf Antisymmetric~~state:}~~~~~\delta E^{N}_{A}=\langle A|H_{LS}|A\rangle =-\frac{i}{2}\sum^{N}_{\delta,\eta=1}\sum^{3}_{i,j=1}H^{(\delta\eta)}_{ij} \langle A|(n^{\delta}_{i}.\sigma^{\delta}_{i})(n^{\eta}_{j}.\sigma^{\eta}_{j})| A\rangle=\frac{P{\cal F}(L,k,\omega_0)\Gamma^{N}_{3;{\cal DC}}}{{\cal N}^2_{\rm norm}}.~~~~~~~~~
\eea
\end{widetext}
Here the overall normalisation factor is appearing from the $N$ entangled spin states, which is given by, ${\cal N}_{\rm norm}=1/\sqrt{{}^{N}C_2}=\sqrt{2(N-2)!/N!}$. For the computation of the matrix elements in the above mentioned shifts we have used the following results:
\begin{widetext}
\bea  \sum^{N}_{\delta,\eta=1}\sum^{3}_{i,j=1}\langle G|(n^{\delta}_{i}.\sigma^{\delta}_{i})(n^{\eta}_{j}.\sigma^{\eta}_{j})| G\rangle &=&\frac{1}{{\cal N}^2_{\rm norm}}\underbrace{\sum^{N}_{\delta,\eta=1}\sum^{N}_{\delta^{'},\eta^{'}=1,\delta^{'}<\eta^{'}}\sum^{N}_{\delta^{''},\eta^{''}=1,\delta^{''}<\eta^{''}}\sum^{3}_{i,j=1}\langle g_{\eta^{'}}|\otimes \langle g_{\delta^{'}}|(n^{\delta}_{i}.\sigma^{\delta}_{i})(n^{\eta}_{j}.\sigma^{\eta}_{j})| g_{\delta^{''}}\rangle \otimes|g_{\eta^{''}}\rangle}_{\textcolor{red}{\equiv ~\Gamma^{N}_{1;{\cal DC}}}}\nonumber\\
&=&\frac{1}{{\cal N}^2_{\rm norm}}\Gamma^{N}_{1;{\cal DC}},~~~~~~~\\
\sum^{N}_{\delta,\eta=1}\sum^{3}_{i,j=1}\langle E|(n^{\delta}_{i}.\sigma^{\delta}_{i})(n^{\eta}_{j}.\sigma^{\eta}_{j})| E\rangle &=&\frac{1}{{\cal N}^2_{\rm norm}}\underbrace{\sum^{N}_{\delta,\eta=1}\sum^{N}_{\delta^{'},\eta^{'}=1,\delta^{'}<\eta^{'}}\sum^{N}_{\delta^{''},\eta^{''}=1,\delta^{''}<\eta^{''}}\sum^{3}_{i,j=1}\langle e_{\eta^{'}}|\otimes \langle e_{\delta^{'}}|(n^{\delta}_{i}.\sigma^{\delta}_{i})(n^{\eta}_{j}.\sigma^{\eta}_{j})| e_{\delta^{''}}\rangle \otimes |e_{\eta^{''}}\rangle}_{\textcolor{red}{\equiv~ \Gamma^{N}_{1;{\cal DC}}}}\nonumber\\
&=&\frac{1}{{\cal N}^2_{\rm norm}}\Gamma^{N}_{1;{\cal DC}},\eea
\bea
&& \sum^{N}_{\delta,\eta=1}\sum^{3}_{i,j=1}\langle S|(n^{\delta}_{i}.\sigma^{\delta}_{i})(n^{\eta}_{j}.\sigma^{\eta}_{j})| S\rangle \nonumber\\
&=&\frac{1}{2{\cal N}^2_{\rm norm}}\underbrace{\sum^{N}_{\delta,\eta=1}\sum^{N}_{\delta^{'},\eta^{'}=1,\delta^{'}<\eta^{'}}\sum^{N}_{\delta^{''},\eta^{''}=1,\delta^{''}<\eta^{''}}\sum^{3}_{i,j=1}(\langle e_{\eta^{'}}\rangle \otimes \langle g_{\delta^{'}}|+\langle g_{\eta^{'}}| \otimes \langle e_{\delta^{'}}|)|(n^{\delta}_{i}.\sigma^{\delta}_{i})(n^{\eta}_{j}.\sigma^{\eta}_{j})|(| e_{\delta^{''}}\rangle \otimes|g_{\eta^{''}}\rangle+| g_{\delta^{''}}\rangle \otimes |e_{\eta^{''}}\rangle)}_{\textcolor{red}{\equiv~ \Gamma^{N}_{2;{\cal DC}}}}\nonumber\\
&=&-\frac{1}{2{\cal N}^2_{\rm norm}}\Gamma^{N}_{2;{\cal DC}},\eea
\bea
&& \sum^{N}_{\delta,\eta=1}\sum^{3}_{i,j=1}\langle A|(n^{\delta}_{i}.\sigma^{\delta}_{i})(n^{\eta}_{j}.\sigma^{\eta}_{j})| A\rangle \nonumber\\
&=&\frac{1}{2{\cal N}^2_{\rm norm}}\underbrace{\sum^{N}_{\delta,\eta=1}\sum^{N}_{\delta^{'},\eta^{'}=1,\delta^{'}<\eta^{'}}\sum^{N}_{\delta^{''},\eta^{''}=1,\delta^{''}<\eta^{''}}\sum^{3}_{i,j=1}(\langle e_{\eta^{'}}\rangle \otimes \langle g_{\delta^{'}}|-\langle g_{\eta^{'}}| \otimes \langle e_{\delta^{'}}|)|(n^{\delta}_{i}.\sigma^{\delta}_{i})(n^{\eta}_{j}.\sigma^{\eta}_{j})|(| e_{\delta^{''}}\rangle \otimes|g_{\eta^{''}}\rangle-| g_{\delta^{''}}\rangle \otimes |e_{\eta^{''}}\rangle)}_{\textcolor{red}{\equiv~ \Gamma^{N}_{3;{\cal DC}}}}\nonumber\\
&=&-\frac{1}{2{\cal N}^2_{\rm norm}}\Gamma^{N}_{3;{\cal DC}}.
\eea
\end{widetext}
Here we found from our computation that the direction cosine dependent factors which are coming as an outcome of the $\underbrace{\cdots}$ highlighted contributions are exactly same for ground and excited states, so that the shifts are also appearing to be exactly same with same signature. On the other hand, from the symmetric and antisymmetric states we have found that he direction cosine dependent highlighted factors are not same. Consequently, the shifts are not also same for these two states. Now one can fix the principal value of the Hilbert transformed integral of the Wightman functions to be unity ($P=1$) for the sake of simplicity, as it just serves the purpose of a overall constant scaling of the computed shifts from all the entangled states for $N$ spins. The explicit expressions for these direction cosine dependent factors are extremely complicated to write for any general large value of the number of $N$ spins. For this reason we have not presented these expressions explicitly in this paper. However, for $N=2$ and $N=3$ spin systems we have presented the results just in the previous section of this supplementary material of this paper. Finally, one can write the following expression for the ratio of the spectroscopic shifts with the corresponding direction cosine dependent factor in a compact notation is derived as:
\begin{eqnarray}
\frac{\delta E_{Y}^{N}}{2\Gamma_{1;{\cal DC}}^N}=\frac{\delta E_{S}^{N}}{\Gamma_{2;{\cal DC}}^N}=-\frac{\delta E_{A}^{N} }{\Gamma_{3;{\cal DC}}^N}=-{\cal F}(L,\omega_0,k)/{\cal N}^2_{\rm norm}, ~~~~~~~
\end{eqnarray} 
where $Y$ represents the ground and the excited states and $S$ and $A$ 
symmetric and antisymmetric states, respectively. Here, $\Gamma^N_{i;{\cal DC}}~\forall~i=1,2,3$ represent the direction cosine dependent angular factor which appears due to the fact that we have considered any arbitrary orientation of $N$ number of identical spins. This result explicitly shows that the ratio of all these shifts with their corresponding direction cosine dependent factor proportional to a spectral function ${\cal F}(L,\omega_0,k)$, given by,
\begin{eqnarray}\label{sdsd}
{\cal F}(L,k,\omega_0)&=&{\cal E}(L,k)\cos\left(2\omega_0 k\sinh^{-1}\left(L/2k\right)\right),~~~
\end{eqnarray}
where, $${\cal E}(L,k)=\mu^2/(8\pi L\sqrt{1+(L/2k)^2}).$$ Here this spectral function is very important as it is the only contribution in this computation which actually directly captures the contribution of the static patch of the de Sitter space-time through the parameter $k$. In this computation we are dealing with two crucial length scale which are both appearing in the spectral function ${\cal F}(L,k,\omega_0)$, which are:
\begin{enumerate}
\item Euclidean distance $L$ and
\item Parameter $k$ which plays the role of  inverse curvature in this problem. 
\end{enumerate}
Depending on these two length scales to analyse the behaviour of this spectral function we have considered two limiting situations, which are given by:
\begin{itemize}
\item \underline{Region $L\gg k$}, which is very useful for our computation as it captures the effect of both the length scale $L$ and $k$. We have found that to determine the observed value of the Cosmological Constant at the present day in Planckian unit this region gives very important contribution.
\item \underline{Region $L\ll k$}, which replicates the analogous effect of Minkowski flat space-time in the computation of spectral shifts. This limiting result may not be very useful for our computation, but clearly shows that exactly when we will loose all the information of the static patch of the de Sitter space. For this reason this region is also not useful at all to determine the value of the observationally consistent value of Cosmological Constant from the spectral shifts. In the later section of this supplementary material it will be shown that if we start doing the same computation of spectral shifts in exactly Minkowski flat space-time then we will get the same results of the spectral shifts that we have obtained in this limiting region.
\end{itemize}
In different euclidean length scales, we have the following approximated expressions for the above mentioned function:
\begin{widetext}
\begingroup
\large
\begin{eqnarray}
	{\cal F}(L,k,\omega_0)=\large
	\left\{
	\begin{array}{lr}
		\displaystyle \frac{\mu^2 k}{4\pi L^2}\cos\left(2\omega_0 k\ln\left(L/2k\right)\right), & \text{$L>>k$}\\ \\
		\displaystyle  \frac{\mu^2}{8\pi L}\cos\left(\omega_0 L\right). & \text{$L<< k$}
	\end{array}
	\right.~~~~~
\end{eqnarray}
\endgroup
\end{widetext} 
{\section{F. Large $N$ limit of spectroscopic shifts}
\label{Appendix:F}}
In this section our objective is to derive the expression for shifts at large $N$ limit. This large $N$ limit is very useful to describe a realistic system in nature and usually identified to be the thermodynamic limit. Stirling's approximation is very useful to deal with factorials of very large number. The prime reason of using Stirling's approximation is to estimate a correct numerical value of the factorial of very large number, provided small error will appear in this computation. However, this is really useful as numerically dealing with the factorial of very large number is extremely complicated job to perform and in some cases completely impossible to perform. In our computation this large number is explicitly appearing in the normalization constant of the entangled states, ${\cal N}_{\rm norm}=1/\sqrt{{}^{N}C_2}=\sqrt{2(N-2)!/N!}$, which we will further analytically estimate using Stirling's formula. Now, according to this approximation one can write the expression for the factorial of a very large number (in our context that number $N$ correspond to the number of spins) as:
\begin{widetext}
\bea \textcolor{blue}{\bf Stirling's ~formula:}~~~~~~N!\sim \sqrt{2N\pi}~\left(\frac{N}{e}\right)^{N}~\left(1+\underbrace{\frac{1}{12N}+{\cal O}\left(\frac{1}{N^2}\right)+\cdots}_{\textcolor{red}{small~corrections}}\right),~~~~~~~~\eea
\end{widetext}
which finally leads to the following bound on $N!$, where $N$ is a positive integer for our system, as:
\begin{widetext}
\bea && \sqrt{2\pi}~N^{N+\frac{1}{2}}~\exp(-N)~\exp\left(\frac{1}{12N+1}\right)\leq N! \leq \exp(1)~N^{N+\frac{1}{2}}~\exp(-N)~\exp\left(\frac{1}{12N}\right).~~~~~~~~\eea
\end{widetext}
Later Gosper had introduced further modification in the Stirling's formula to get more accurate answer of the factorial of a very large number, which is given by the following expression:
\begin{widetext}
\bea \textcolor{blue}{\bf Stirling~Gosper ~formula:}~~~~~~N!\sim \sqrt{\left(2N+\underbrace{\frac{1}{3}}_{\textcolor{red}{Gosper~factor}}\right)\pi}~\left(\frac{N}{e}\right)^{N}~\left(1+\underbrace{\frac{1}{12N}+{\cal O}\left(\frac{1}{N^2}\right)+\cdots}_{\textcolor{red}{small~corrections}}\right).~~~~~~~~\eea
\end{widetext}
Using this formula one can further evaluate the expression for $(N-2)!$ for $N$ spin system as:
\begin{widetext}
\bea (N-2)!\sim \sqrt{\left(2N-\frac{11}{3}\right)\pi}~\left(\frac{N-2}{e}\right)^{N-2}~\left(1+\underbrace{\frac{1}{12(N-2)}+{\cal O}\left(\frac{1}{(N-2)^2}\right)+\cdots}_{\textcolor{red}{small~corrections}}\right).~~~~~~~~\eea
\end{widetext}
Here we want to point out few more revised version of the Stirling's formula, which are commonly used in various contexts: \cite{gamma1, MORTICI20111351}
\begin{widetext}
\bea &&\textcolor{blue}{\bf Stirling~Burnside~ formula:}~~~~~~N!\sim\sqrt{2\pi}\left(\frac{N+\frac{1}{2}}{e}\right)^{N+\frac{1}{2}},\\
&&\textcolor{blue}{\bf Stirling~Ramanujan~ formula:}~~~~~~N!\sim\sqrt{2\pi}\left(\frac{N}{e}\right)^{N}\left(N^3+\frac{1}{2}N^2+\frac{1}{8}N+\frac{1}{240}\right)^{1/6},\\ 
&&\textcolor{blue}{\bf Stirling~Windschitl~ formula:}~~~~~~N!\sim\sqrt{2\pi N}\left(\frac{N}{e}\right)^{N}\left(N\sinh \frac{1}{N}\right)^{N/2},\\
&&\textcolor{blue}{\bf Stirling~Nemes~ formula:}~~~~~~N!\sim\sqrt{2\pi N}\left(\frac{N}{e}\right)^{N}\left(1+\frac{1}{12N^2-\frac{1}{10}}\right)^{N}. \eea
\end{widetext}
Further in the large $N$ limit, using the {\it Stirling-Gosper} approximation, the normalization factor can be written as:
\begin{widetext}
\begin{eqnarray}
	&&{\cal N}_{\rm norm}=\frac{1}{\sqrt{{}^{N}C_2}}~~ \underrightarrow{\rm Large~N}~~\widehat{{\cal N}_{\rm norm}} \approx\sqrt{2}\left(1-\frac{2}{\left(N+\frac{1}{6}\right)}\right)^{1/4}\left(\frac{N}{e}\right)^{-N/2}\left(\frac{N-2}{e}\right)^{N/2-1}\sqrt{\frac{1-\frac{2}{\left(N+\frac{1}{12}\right)}}{\left(1-\frac{2}{N}\right)}}.~~~~ \end{eqnarray}
\end{widetext}
Thus in the large $N$ limit the spectral shifts can be approximately derived as :
\begin{eqnarray}
\frac{\widehat{\delta E_{Y}^{N}}}{2\Gamma_{1;{\cal DC}}^N}=\frac{\widehat{\delta E_{S}^{N}}}{\Gamma_{2;{\cal DC}}^N}=-\frac{\widehat{\delta E_{A}^{N}} }{\Gamma_{3;{\cal DC}}^N}=-{\cal F}(L,k,\omega_0) /\widehat{{\cal N}_{\rm norm}}^2. ~~~~~~~~
\end{eqnarray} 
From the above expressions derived in the large $N$ limit we get the following information: 
\begin{itemize}
\item Contribution from the large $N$ limit will only effect the normalization factors appearing in the shifts,

\item The prime contribution, which is comping from the spectral function ${\cal F}(L,\omega_0,k)$ is independent of the number $N$. So it is expected that directly this contribution will not be effected by the large $N$ limiting approximation in the factorial.
\end{itemize}
{\section{G. Flat space limit of spectroscopic shifts for $N$ spins}
\label{Appendix:G}}
Now, our objective is to the obtained results for spectroscopic shifts in the $L<<k$ limit with the result one can derive in the context of the Minkowski flat space. Considering the same physical set up, the two point thermal correlation functions can be expressed in terms of the $N$ spin Wightman function for massless probe scalar field can be expressed as:
\begin{widetext}
\begin{eqnarray}
	G^{\rm Min}_{N}(x,x^{'})&=&\begin{pmatrix} ~
		\underbrace{G^{\delta\delta}_{\rm Min}(x,x')}_{\textcolor{red}{Auto-Correlation}}~~~ &~~~ \underbrace{G^{\delta\eta}_{\rm Min}(x,x')}_{\textcolor{red}{Cross-Correlation}}~ \\
		~\underbrace{G^{\eta\delta}_{\rm Min}(x,x')}_{\textcolor{red}{Cross-Correlation}}~~~ &~~~ \underbrace{G^{\eta\eta}_{\rm Min}(x,x') }_{\textcolor{red}{Auto-Correlation}}~
	\end{pmatrix}_{\beta}=\begin{pmatrix} ~
		\langle \hat{\Phi}({\bf x_{\delta}},\tau)\Phi({\bf x_{\delta}},\tau')\rangle_{\beta}~~~ &~~~ \langle \hat{\Phi}({\bf x_{\delta}},\tau)\Phi({\bf x_{\eta}},\tau')\rangle_{\beta}~ \\ \\ \\
		~\langle \hat{\Phi}({\bf x_{\eta}},\tau)\Phi({\bf x_{\delta}},\tau')\rangle_{\beta}~~~ &~~~ \langle \hat{\Phi}({\bf x_{\eta}},\tau)\Phi({\bf x_{\eta}},\tau')\rangle_{\beta} ~
	\end{pmatrix}_{\rm Min}~,\nonumber\\
	&& ~~~~~~~~~~~~~~~~~~~~~~~~~~~~~~~~~~~~~~~~~~~~~~~~~~~~~~~~~~~~~\forall~~\delta,\eta=1,\cdots,N~{\rm (for~both~even~\&~ odd)}.
\end{eqnarray} 
\end{widetext}
where the individual Wightman functions can be computed using the well known {\it Schwinger Keldysh} path integral technique as:
\begin{widetext}
\bea G^{\delta\delta}_{\rm Min}(x,x')&=&-\frac{1}{4\pi^2}\sum^{\infty}_{m=-\infty}\frac{1}{\left(\Delta\tau-i\left\{2\pi k m+\epsilon\right\}\right)^2}=\frac{1}{16 \pi ^2 k^2}~{\rm cosec}^2\left(\frac{\epsilon+i \Delta\tau}{2 k}\right),\\
G^{\delta\eta}_{\rm Min}(x,x')
&=&-\frac{1}{4\pi^2}\sum^{\infty}_{m=-\infty}\frac{1}{\left(\Delta\tau-i\left\{2\pi k m+\epsilon\right\}\right)^2-L^2}=\frac{1}{16 \pi ^2 k L}~
\left[2 \left\{{\rm Floor}\left( \frac{1}{2 \pi }\arg \left(\frac{\epsilon+i (\Delta\tau+L)}{k}\right)\right)\right.\right.\nonumber\\
&& \left.\left.~~~~~~~~~~~~~~~~~~~~~~~~~~~~~~~~~~~~~~~~~~~~~~~~~~~~~~~~~~~~~~~~~~~~~~~~~~~~~~~~~~-{\rm Floor}\left( \frac{1}{2\pi}\arg \left(\frac{\epsilon+i (\Delta\tau-L)}{k}\right)\right)\right\} \right.\nonumber\\
&& \left.~~~~~~~~~~~~~~~~~~~~~~~~~~~~~~~~~~~~~~~~~~~~~~~~~~~~~~~~~~~~~~ +i \left\{\cot \left(\frac{\epsilon+i (\Delta\tau+L)}{2 k}\right)-\cot \left(\frac{\epsilon+i (\Delta\tau-L)}{2 k}\right)\right\}\right],~~~~~~~
\eea
\end{widetext}
where $\epsilon$ is an infinitesimal quantity which is introduced to deform the contour of the path integration.  Using this Wightman function we can carry forward the similar calculation for spectroscopic shifts in Minkowsi space, which gives:
\begin{widetext}
\begin{eqnarray}
	&&\textcolor{blue}{\bf For~ general~ N:}~~~~~
	\underbrace{\frac{\delta E_{Y,{\rm Min}}^{N}}{2\Gamma_{1;{\cal DC}}^N}=\frac{\delta E_{S,{\rm Min}}^{N}}{\Gamma_{2;{\cal DC}}^N}=-\frac{\delta E_{A,{\rm Min}}^{N} }{\Gamma_{3;{\cal DC}}^N}}_{\textcolor{red}{Minkowski~space~calculation}}=-\cos\left(\omega_0 L\right)  /{\cal N}^2_{\rm norm}=\underbrace{\frac{\delta E_{Y,{\rm Min}}^{N}}{2\Gamma_{1;{\cal DC}}^N}=\frac{\delta E_{S,{\rm Min}}^{N}}{\Gamma_{2;{\cal DC}}^N}=-\frac{\delta E_{A,{\rm Min}}^{N} }{\Gamma_{3;{\cal DC}}^N}}_{\textcolor{red}{Region~L\ll k~calculation}}, ~~~~~~~\\
	&&\textcolor{blue}{\bf For~ large~ N:}~~~~~
	\underbrace{\frac{\widehat{\delta E_{Y,{\rm Min}}^{N}}}{2\Gamma_{1;{\cal DC}}^N}=\frac{\widehat{\delta E_{S,{\rm Min}}^{N}}}{\Gamma_{2;{\cal DC}}^N}=-\frac{\widehat{\delta E_{A,{\rm Min}}^{N}} }{\Gamma_{3;{\cal DC}}^N}}_{\textcolor{red}{Minkowski~space~calculation}}=-\cos\left(\omega_0 L\right)  /\widehat{{\cal N}_{\rm norm}}^2=\underbrace{\frac{\widehat{\delta E_{Y,{\rm Min}}^{N}}}{2\Gamma_{1;{\cal DC}}^N}=\frac{\widehat{\delta E_{S,{\rm Min}}^{N}}}{\Gamma_{2;{\cal DC}}^N}=-\frac{\widehat{\delta E_{A,{\rm Min}}^{N}} }{\Gamma_{3;{\cal DC}}^N}}_{\textcolor{red}{Region~L\ll k~calculation}}, 
\end{eqnarray} 
\end{widetext}
where $Y$ represents the ground and the excited states and $S$ and $A$ 
symmetric and antisymmetric states, respectively. Here, $\Gamma^N_{i;{\cal DC}}~\forall~i=1,2,3$ represent the direction cosine dependent angular factor which appears due to the fact that we have considered any arbitrary orientation of $N$ number of identical spins. Here all the quantities in $\widehat{}$ are evaluated at the large $N$ limit by using Stirling Gosper formula as mentioned earlier. Here it is clearly observed that the shifts are independent of the temperature of the thermal bath, $T=1/2\pi k$ and only depends on direction cosines and the euclidean distance $L$. Also we found that this result exactly matches with the result obtained for the limiting case $L\ll k$.

{\section{H. Derivation of the bath Hamiltonian in static patch of de Sitter}
\label{Appendix:H}}

Below we provide the derivation of the bath Hamiltonian. 
The bath is described by a massless probe scalar field, which is given by the following action:
\begin{align} 
S_{\bf Bath}&=\frac{1}{2}\int d^4x \sqrt{-g}~g^{\mu\nu}~\left(\partial_{\mu}\Phi(x)\right)\left(\partial_{\nu}\Phi(x)\right) \\
&=\int dt~d^3x~{\cal L}(g^{\mu\nu},g,\partial_{\mu}\Phi),
\end{align}
where ${\cal L}(g^{\mu\nu},g,\partial_{\mu}\Phi)$ is the Lagrangian density in presence of background gravity, which can be explicitly written as:
\begin{align} 
{\cal L}(g^{\mu\nu},g,\partial_{\mu}\Phi(x))=\frac{1}{2}\sqrt{-g}~g^{\mu\nu}~\left(\partial_{\mu}\Phi(x)\right)\left(\partial_{\nu}\Phi(x)\right). 
\end{align}
Here the scalar field is embedded in static patch of the de Sitter space which is described by the following infinitesimal line element:	
\begin{align}
ds^{2}=\left(1-\frac{r^{2}}{\alpha^{2}}\right)dt^{2}-\left(1-\frac{r^{2}}{\alpha^{2}}\right)^{-1}dr^{2}-r^{2}(d\theta^{2}+\sin^{2}\theta d\phi^{2})
\end{align}
where $\alpha=\sqrt{\frac{3}{\Lambda}}>0$.
Here $r=\alpha$ represents the horizon where we have space like singularity in the metric of static de Sitter space time.

The canonically conjugate momentum for this massless probe scalar field is given by the following expression:
\begin{align}
\label{mom1}\Pi_{\Phi}(x)&\equiv \frac{\partial {\cal L}(g^{\mu\nu},g,\partial_{\mu}\Phi(x))}{\partial(\partial_{0}\Phi(x))}\nonumber\\
&=\frac{\partial {\cal L}(g^{\mu\nu},g,\partial_{\mu}\Phi(x))}{\partial\dot{\Phi}(x)}\nonumber\\
&=\sqrt{-g}~g^{00}\dot{\Phi}(x),
\end{align}
and in static patch of de Sitter space we get:
\begin{align} \Pi_{\Phi}(t,r,\theta,\phi)&=r^2\sin\theta~\left(1-\frac{r^{2}}{\alpha^{2}}\right)^{-1}\dot{\Phi}(t,r,\theta,\phi).
\end{align}
Here we have used the fact that:
\begin{align}
\sqrt{-g}=r^2\sin\theta~~~~{\rm and}~~~~~g^{00}=\left(1-\frac{r^{2}}{\alpha^{2}}\right)^{-1}.
\end{align}
From Eq~(\ref{mom1}), one can further write:
\begin{align} \dot{\Phi}(t,r,\theta,\Phi)=\frac{\Pi_{\Phi}(t,r,\theta,\phi)}{r^2\sin\theta}\left(1-\frac{r^{2}}{\alpha^{2}}\right),
\end{align}
which we will use further to compute the expression for the bath Hamiltonian density.

Further, using Legendre transformation the Hamiltonian density in the static patch of the de Sitter space can be written as:
\begin{align}
{\cal H}_{\bf Bath}&=\Pi_{\Phi}(x)\dot{\Phi}(x)-{\cal L}(g^{\mu\nu},g,\partial_{\mu}\Phi(x))\nonumber\\
&=\frac{\Pi^2_{\Phi}(t,r,\theta,\phi)}{r^2\sin\theta}\left(1-\frac{r^{2}}{\alpha^{2}}\right)-{\cal L}(g^{\mu\nu},g,\partial_{\mu}\Phi(x)).
\end{align}
Now, in the static patch of the de Sitter space the Lagrangian density can be explicitly written as:

\begin{widetext}

\begin{align}
	\label{kjq1}{\cal L}(g^{\mu\nu},g,\partial_{\mu}\Phi(x))&=\frac{1}{2}\sqrt{-g}~g^{\mu\nu}~\left(\partial_{\mu}\Phi(x)\right)\left(\partial_{\nu}\Phi(x)\right)\nonumber\\
	&=\frac{1}{2}r^2\sin\theta\left\{\frac{\Pi^2_{\Phi}(t,r,\theta,\phi)}{r^4\sin^2\theta}\left(1-\frac{r^{2}}{\alpha^{2}}\right)-\left(1-\frac{r^{2}}{\alpha^{2}}\right)(\partial_{r}\Phi(t,r,\theta,\phi))^2\right.\nonumber \\ 
	& \left.~~~~~~~~~~~~~~~~~~~~~~~-\frac{1}{r^2}(\partial_{\theta}\Phi(t,r,\theta,\phi))^2-\frac{1}{r^2\sin^2\theta}(\partial_{\phi}\Phi(t,r,\theta,\phi))^2\right\}\nonumber\\
	&=\frac{1}{2}\left\{\frac{\Pi^2_{\Phi}(t,r,\theta,\phi)}{r^2\sin\theta}\left(1-\frac{r^{2}}{\alpha^{2}}\right)-\left(1-\frac{r^{2}}{\alpha^{2}}\right)r^2\sin\theta(\partial_{r}\Phi(t,r,\theta,\phi))^2\right.\nonumber \\ 
	& \left.~~~~~~~~~~~~~~~~~~~~~~~-\sin\theta(\partial_{\theta}\Phi(t,r,\theta,\phi))^2-\frac{1}{\sin\theta}(\partial_{\phi}\Phi(t,r,\theta,\phi))^2\right\}.~~~~~~~~~
\end{align}	

\end{widetext}

Using Eq~(\ref{kjq1}), we get the following simplified expression for the Hamiltonian density in the static patch of de Sitter space:
\begin{widetext}
\begin{align}
	{\cal H}_{\bf Bath}&=\frac{\Pi^2_{\Phi}(t,r,\theta,\phi)}{2r^2\sin\theta}\left(1-\frac{r^{2}}{\alpha^{2}}\right)+\frac{1}{2}\left\{\left(1-\frac{r^{2}}{\alpha^{2}}\right)r^2\sin\theta(\partial_{r}\Phi(t,r,\theta,\phi))^2\right.\nonumber \\ 
	& \left.~~~~~~~~~~~~~~~~~~~~~~~+\sin\theta(\partial_{\theta}\Phi(t,r,\theta,\phi))^2+\frac{1}{\sin\theta}(\partial_{\phi}\Phi(t,r,\theta,\phi))^2\right\}.
\end{align}
\end{widetext}
Now the 3D spatial volume element in the static patch of the de Sitter space is given by the following expression:
\begin{align} d^3x=r^2\sin\theta~\left(1-\frac{r^{2}}{\alpha^{2}}\right)^{-1}~dr~d\theta~d\phi.\end{align}	

Hence using this 3D spatial volume element the Hamiltonian of the bath in the static patch of the de Sitter space is given by the following expression:
\begin{widetext}
\begin{align}
	H_{\bf Bath}&=\int d^3x ~{\cal H}_{\bf Bath}\nonumber\\
	&=\int^{\alpha}_{0} dr~\int ^{\pi}_{0}d\theta~\int^{2\pi}_{0}d\phi~r^2\sin\theta~\left(1-\frac{r^{2}}{\alpha^{2}}\right)^{-1}\nonumber\\
	&~~~~~~~~~~~~~~\times \left[\frac{\Pi^2_{\Phi}(t,r,\theta,\phi)}{2r^2\sin\theta}\left(1-\frac{r^{2}}{\alpha^{2}}\right)+\frac{1}{2}\left\{\left(1-\frac{r^{2}}{\alpha^{2}}\right)r^2\sin\theta(\partial_{r}\Phi(t,r,\theta,\phi))^2\right.\right.\nonumber \\ 
	& \left.\left.~~~~~~~~~~~~~~~~~~~~~~~+\sin\theta(\partial_{\theta}\Phi(t,r,\theta,\phi))^2+\frac{1}{\sin\theta}(\partial_{\phi}\Phi(t,r,\theta,\phi))^2\right\}\right]\nonumber\\
	&=\int^{\alpha}_{0} dr~\int ^{\pi}_{0}d\theta~\int^{2\pi}_{0}d\phi~\left[\frac{\Pi^2_{\Phi}(\tau,r,\theta,\phi)}{2}\right.\nonumber\\
	& \left.~~~~+\frac{r^2\sin^2\theta}{2}\left\{r^2~(\partial_{r}\Phi(\tau,r,\theta,\phi))^2+ \frac{\left((\partial_{\theta}\Phi(\tau,r,\theta,\phi))^2+\frac{(\partial_{\phi}\Phi(\tau,r,\theta,\phi))^2}{\sin^2\theta}\right)}{\displaystyle\left(1-\frac{r^2}{\alpha^2}\right)}\right\}\right].~~~~~~~~~
\end{align}
\end{widetext}

In this description, $r=\alpha$, which is the upper limit of the radial integral physically represents the horizon in static patch of de Sitter space.

Here it is important to note that, if we further take the $\alpha\rightarrow \infty$ limit then we get the following result:
\begin{widetext}
\begin{align}
	H_{\bf Bath}
	&=\int^{\infty}_{0} dr~\int ^{\pi}_{0}d\theta~\int^{2\pi}_{0}d\phi~\left[\frac{\Pi^2_{\Phi}(\tau,r,\theta,\phi)}{2}\right.\nonumber\\
	& \left.~~~~+\frac{r^2\sin^2\theta}{2}\left\{r^2~(\partial_{r}\Phi(\tau,r,\theta,\phi))^2+\left((\partial_{\theta}\Phi(\tau,r,\theta,\phi))^2+\frac{(\partial_{\phi}\Phi(\tau,r,\theta,\phi))^2}{\sin^2\theta}\right)\right\}\right],
\end{align}
\end{widetext}
which represents the Hamiltonian of a sphere with radius $R$.

\clearpage

\end{document}